\RequirePackage{lineno}
\documentclass[twocolumn,prc,aps,showpacs,floatfix,amsmath,amssymb]{revtex4}
\usepackage{graphicx}   
\usepackage{placeins}

\usepackage{xspace}
\usepackage{color}
\RequirePackage{lineno}

\newcommand{\Rout}{\mbox{$R_{\text{out}}$}\xspace}
\newcommand{\Rside}{\mbox{$R_{\text{side}}$}\xspace}
\newcommand{\Rlong}{\mbox{$R_{\text{long}}$}\xspace}
\newcommand{\pt}{\mbox{$p_{\text{T}}$}\xspace}
\newcommand{\mt}{\mbox{$m_{\text{T}}$}\xspace}

\begin{document}




\title{Flow and interferometry results from Au+Au collisions at $\sqrt{s_{\text{NN}}}$ = 4.5 GeV}

\author{
J.~Adam$^{6}$,
L.~Adamczyk$^{2}$,
J.~R.~Adams$^{39}$,
J.~K.~Adkins$^{30}$,
G.~Agakishiev$^{28}$,
M.~M.~Aggarwal$^{41}$,
Z.~Ahammed$^{61}$,
I.~Alekseev$^{3,35}$,
D.~M.~Anderson$^{55}$,
A.~Aparin$^{28}$
% ...
\centerline{(STAR Collaboration)}
}

% ...
\address{$^{2}$AGH University of Science and Technology, FPACS, Cracow 30-059, Poland}
\address{$^{3}$Alikhanov Institute for Theoretical and Experimental Physics NRC "Kurchatov Institute", Moscow 117218, Russia}
% ...
\address{$^{6}$Brookhaven National Laboratory, Upton, New York 11973}
% ...
\address{$^{28}$Joint Institute for Nuclear Research, Dubna 141 980, Russia}
% ...
\address{$^{30}$University of Kentucky, Lexington, Kentucky 40506-0055}
% ...
\address{$^{35}$National Research Nuclear University MEPhI, Moscow 115409, Russia}
% ...
\address{$^{39}$Ohio State University, Columbus, Ohio 43210}
% ...
\address{$^{41}$Panjab University, Chandigarh 160014, India}
% ...
\address{$^{55}$Texas A\&M University, College Station, Texas 77843}
% ...
\address{$^{61}$Variable Energy Cyclotron Centre, Kolkata 700064, India}
% ...

\date{\today}
\begin{abstract}
The beam energy scan (BES) program at the Relativistic Heavy Ion Collider (RHIC) was extended to energies below $\sqrt{s_{\text{NN}}} = 7.7$ GeV in 2015 by successful implementation of the fixed-target mode of operation in the STAR (Solenoidal Tracker At RHIC) experiment.  In this mode, ions circulate in one ring of the collider and interact with a stationary target at the entrance of the STAR Time Projection Chamber.  The first results for Au+Au collisions at $\sqrt{s_{\text{NN}}} = 4.5$ GeV are presented, demonstrating good performance of all the relevant detector subsystems in fixed-target mode. Results presented here include directed and elliptic flow of identified hadrons, and radii from pion femtoscopy. The latter, together with recent HADES results, reveal a long-sought peak structure that may be caused by the system evolving through a first-order phase transition from quark-gluon plasma to the hadronic phase. Directed and elliptic flow for pions are presented for the first time at this beam energy. Pion and proton elliptic flow show behavior which hints at constituent quark scaling, and demonstrate that a definitive conclusion will be achievable using the full statistics of the on-going second phase of BES (BES-II). In particular, BES-II to date has recorded fixed-target data sets with two orders of magnitude more events at each of nine energies between $\sqrt{s_{\text{NN}}} = 3.0$ and 7.7 GeV.  

\end{abstract}


\maketitle

\section{Introduction}
The beam energy scan (BES) program at RHIC was undertaken to study the nature of the 
Quantum Chromodynamics (QCD) phase diagram in the plane of temperature versus baryon chemical potential, which is explored by varying the collision energy when heavy nuclei interact \cite{Rischke:2003mt, Stephanov:2004wx, Fukushima:2013rx, Brambilla:2014jmp}. The phase diagram region of current interest, at relatively high baryon chemical potential, is not accessible so far by first-principle lattice QCD calculations \cite{Bazavov_2020}. There is thus a wide-ranging international effort to investigate it experimentally \cite{Galatyuk:2019lcf}.

The BES-II program covers collision energies at and below $\sqrt{s_{\rm NN}} = 19.6$ GeV and has the goals of investigating the turn-off of the quark-gluon plasma (QGP) signatures already reported at higher beam energies, and of searching for evidence for a possible first-order phase transition and a critical point \cite{starcollaboration2010experimental, STAR:SN0598}.  The lowest beam energy which is accessible at RHIC with adequate luminosity in the collider mode of operation is $\sqrt{s_{\rm NN}} = 7.7$ GeV.  Therefore a fixed-target (FXT) program has been developed to broaden the reach of BES-II and allow the STAR experiment \cite{ACKERMANN2003624} to access energies below $\sqrt{s_{\rm NN}} = 7.7$ GeV. In this paper, results are presented from a first run using a single RHIC beam at the normal injection energy ($E_{\rm total}$ = 9.8 GeV/nucleon, $E_{\rm kinetic}$ = 8.9 GeV/nucleon) incident on a gold target inside STAR beam-pipe, providing Au+Au collisions at $\sqrt{s_{\rm NN}} = 4.5$ GeV.  In reporting a small subsample of data at a single beam energy, we address a subset of the BES-II goals; moreover, the current results have broad implications by virtue of being the first demonstration of STAR's capability to use FXT mode to extend studies lower in beam energy than previously possible.  

Similar Au+Au collision energies were studied during the fixed-target heavy-ion program at the Alternating Gradient Synchrotron (AGS) in the 1990s \cite{RAI1999162}, covering the range 2.7 to 4.9 GeV in $\sqrt{s_{\rm NN}}$.  The present measurements of heavy-ion collisions at 4.5 GeV with STAR in FXT mode extend the systematics of the world data on a number of observables at these energies. Note that the AGS/E895 measurements are the only available data for a heavy system (Au+Au) near $\sqrt{s_{\rm NN}} = 4.5$ GeV. The CERN energy scan by the NA49 experiment with Pb+Pb collisions reported data at higher energies, namely at $\sqrt{s_{\rm NN}} = 6.4$ GeV and above.

\section{Experimental Setup}

\begin{figure}[ht]
\includegraphics[width=0.48\textwidth]{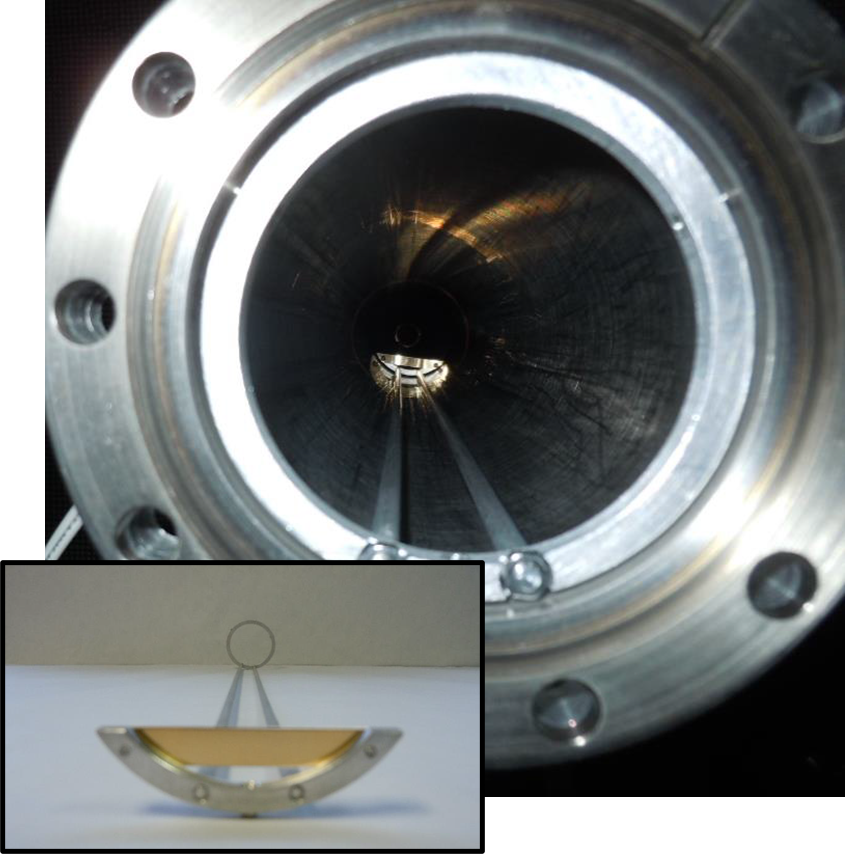}
\caption{\label{fig:Gold_Target} (Color online) Photo of the gold target inserted inside the beam pipe. Inset: photo of the gold target on its aluminum support structure.}
\end{figure}

\begin{figure}[ht]
\includegraphics[width=0.48\textwidth]{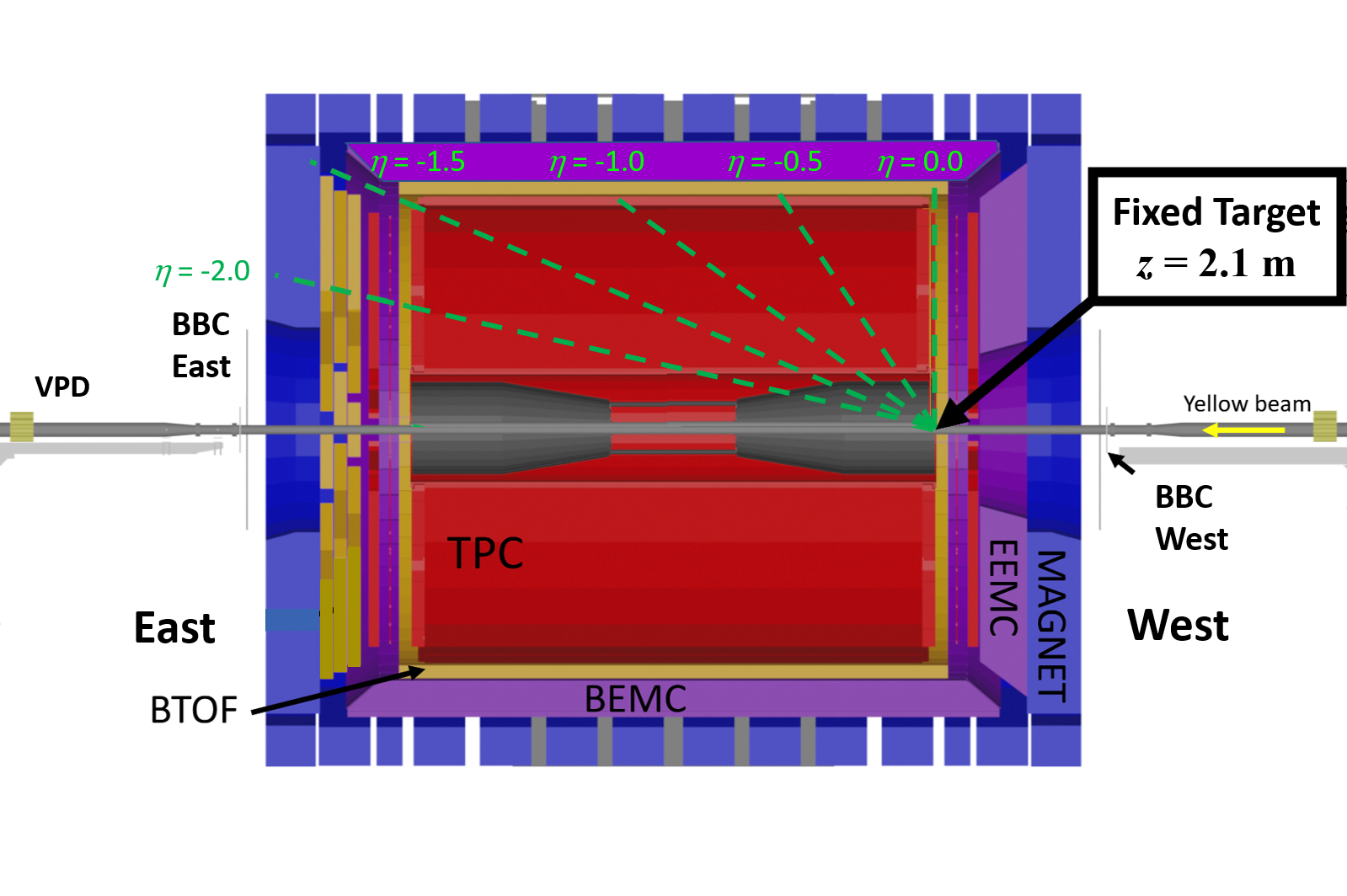}
\caption{\label{fig:Experimental_Setup} (Color online) A schematic cross section of the STAR detector, showing the location of the target.}
\end{figure}

For the results reported in this paper, RHIC provided a single beam of gold ions with a kinetic energy of 8.9~GeV per nucleon    
in the laboratory frame.
The beam was incident on a gold target of thickness 1.93 g/cm$^2$ (1 mm), corresponding to a 4\% interaction probability (determined using the inelastic Au+Au cross section).   
The target was installed inside the vacuum pipe, below its center and 211 cm (see Fig.~\ref{fig:Gold_Target}) to the west of the center of the STAR detector (see Fig.~\ref{fig:Experimental_Setup}). RHIC was setup up circulating six bunches with a total beam intensity of $3.4 \times 10^9$ ions, and filled bunches passed the target at a rate of 500 kHz. The beam was then carefully lowered 1.8 cm (note that the radius of the beam pipe in the interior of the detector was 2.0 cm) such that its halo was grazing the top edge of the target. An average of 0.2 gold ions were incident on the target with each passing beam bunch. The target thickness was such that 4\% of the incident gold ions experienced an inelastic hadronic collision. The trigger rate of 1 kHz was influenced by the bunch rate, the number of incident ions per bunch, the interaction probability, and the trigger bias (discussed later). The number of filled bunches was selected to ensure that tracks from out-of-time collisions would not be associated with triggered events in the gold target. The amount of circulating beam allowed to be incident on the target was adjusted to fill the STAR data acquisition bandwidth, while minimizing radiation on the inner silicon detectors (which were not used for this test run). The store was held for one hour, and there was no perceptible loss of beam intensity over that period. The one-hour duration was determined by the time allocated to the proof-of-principle test run. The detector systems used for this test run were the Time Projection Chamber (TPC)~\cite{ANDERSON2003659}, the time-of-flight (TOF)~\cite{Llope:2012zz}, and the Beam-Beam Counter (BBC)~\cite{Judd:2018}. In this fixed-target configuration, the TPC covered a range of polar angles specified by $0.1 < \eta_{\rm lab} < 2$, the TOF covered the range $0.1 < \eta_{\rm lab} < 1.5$, and the BBC, which was only used for triggering, covered the range 
$3.3 < \eta_{\rm lab} < 5.0$.  
This FXT configuration provided tracking and particle identification from target rapidity to midrapidity.  Details of the pion and proton acceptance in rapidity and transverse momentum are shown in the next section.

Central Au+Au events were recorded by requiring a coincidence between the downstream trigger detector, an arrangement of scintillator tiles called the Beam-Beam Counter (BBC) \cite{Judd:2018}, and a high multiplicity signal in the time-of-flight (TOF) barrel \cite{Llope:2012zz}. The TOF multiplicity requirement was 130 or more for the bulk of the data to ensure that the trigger would not fire on collisions between beam halo and the aluminum beam pipe or target support structure. 
Previous studies of collisions between the beam halo and the beam pipe had recorded central Au+Al events with TOF multiplicities as high as 120 tracks. Analysis of the data from this test run indicates that the background was negligible, and that finding has allowed the FXT physics runs performed in 2018, 2019 and 2020 to use minimum-bias triggers. From this brief test run,
about 1.3 million events with centrality 0-30\% were recorded.

\section{Performance in Fixed Target Mode}

\begin{figure}[ht]
\includegraphics[width=0.48\textwidth]{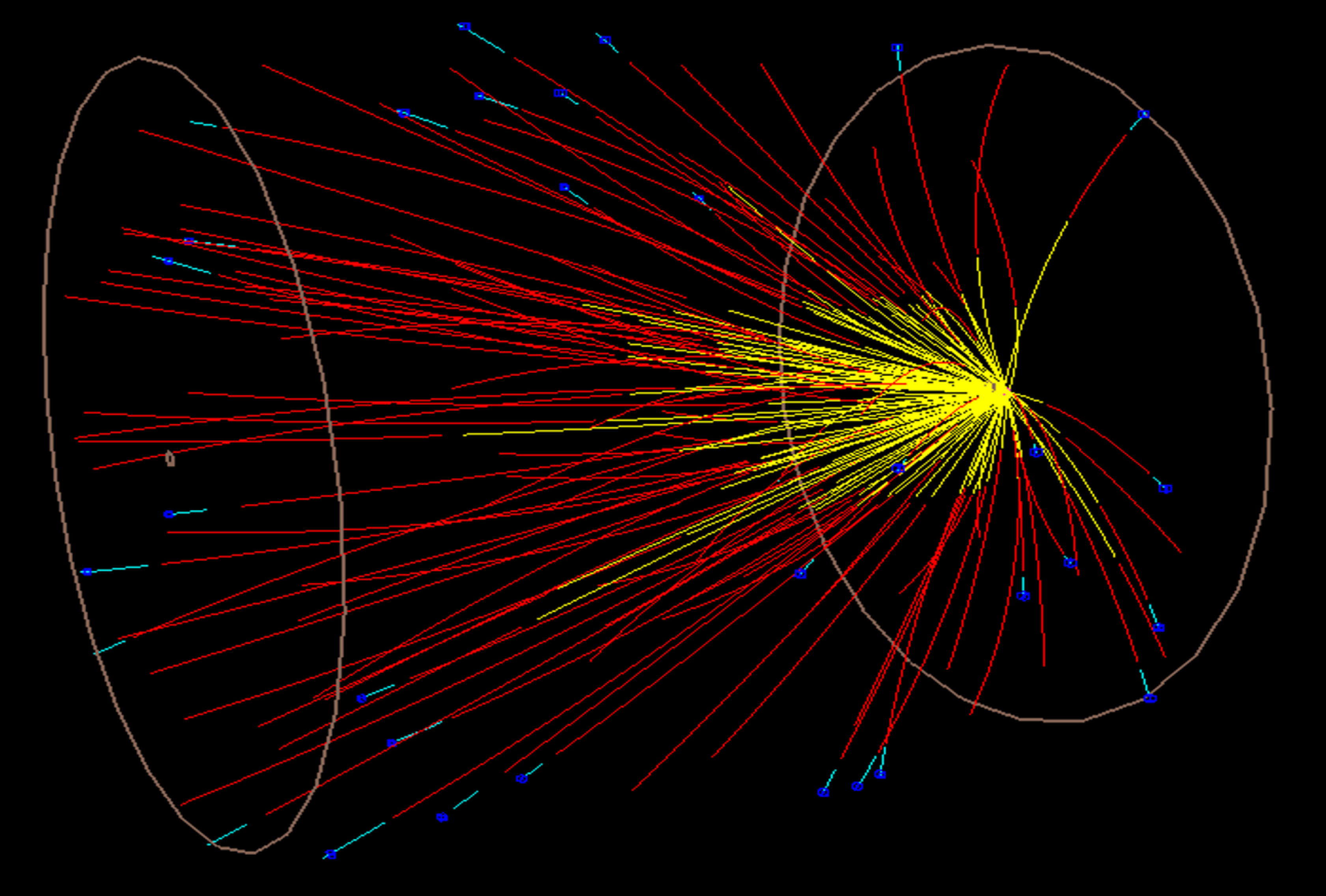}
\caption{\label{fig:AUAU_FXT_Event} (Color online) Reconstruction of a $\sqrt{s_{\mathrm{NN}}}$ = 4.5 GeV Au+Au event. TPC tracks are shown in red, projections to the vertex within the target are shown in yellow, and associated TOF hits are shown in blue.}
\end{figure}

\begin{figure}[ht]
\includegraphics[width=0.48\textwidth]{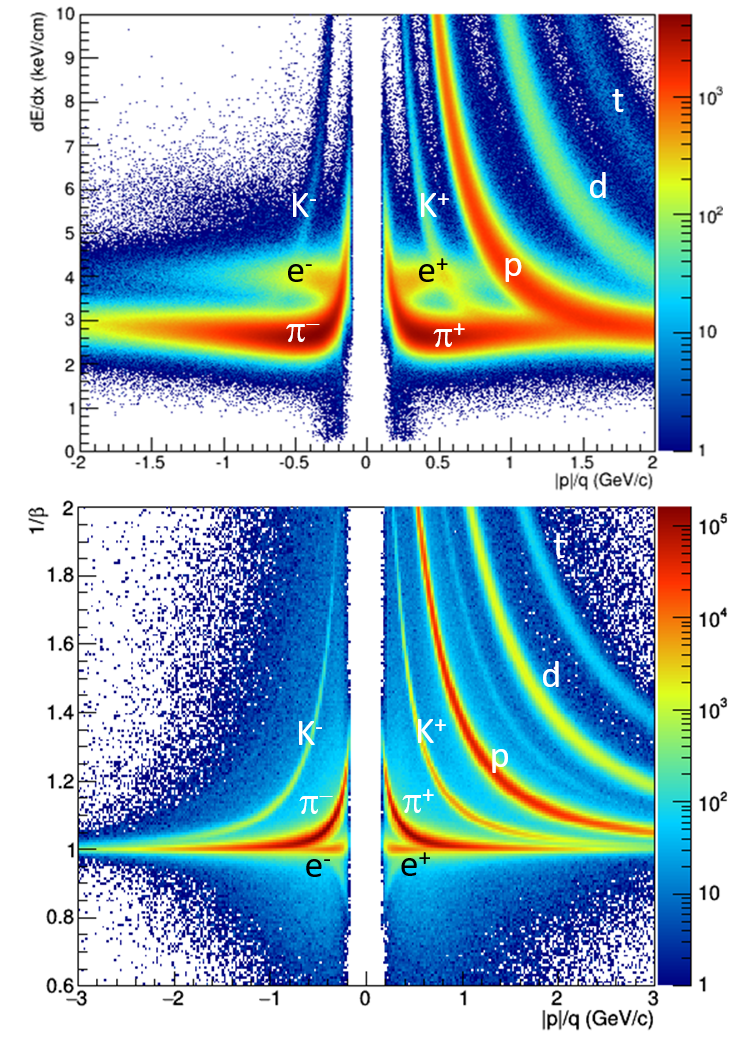}
\caption{\label{fig:PID_map} (Color online) The top panel shows the particle identification using $dE/dx$ in the TPC. The bottom panel shows particle identification using inverse velocity ($1/\beta$) measured by the TOF.}
\end{figure}

As a first indicator of the performance of the STAR detector in fixed-target mode, a reconstructed event is shown in Fig.~\ref{fig:AUAU_FXT_Event}. In some ways, the performance for mid-rapidity tracks in FXT mode exceeds the performance in collider mode. The mid-rapidity tracks are five meters long as opposed to two meters, which improves the TPC $dE/dx$ resolution from 6.8\% to 4.6\%. Furthermore, TOF $K/\pi$ separation is maintained up to 2.5 GeV/$c$ instead of up to 1.6 GeV/$c$ (see Fig.~\ref{fig:PID_map}). 
The lower particle multiplicities in the FXT events compared to those in higher-energy collider mode collisions result in larger tracking efficiencies.
In other ways, the performance for FXT is more challenging. The rapidity boost of the center-of-mass means that a larger fraction of the mid-rapidity particles require TOF hits for particle identification, and the $\eta$ acceptance limits raise the low-$p_T$ cut-offs for kaons and protons.

\begin{figure}[ht]
\includegraphics[width=0.48\textwidth]{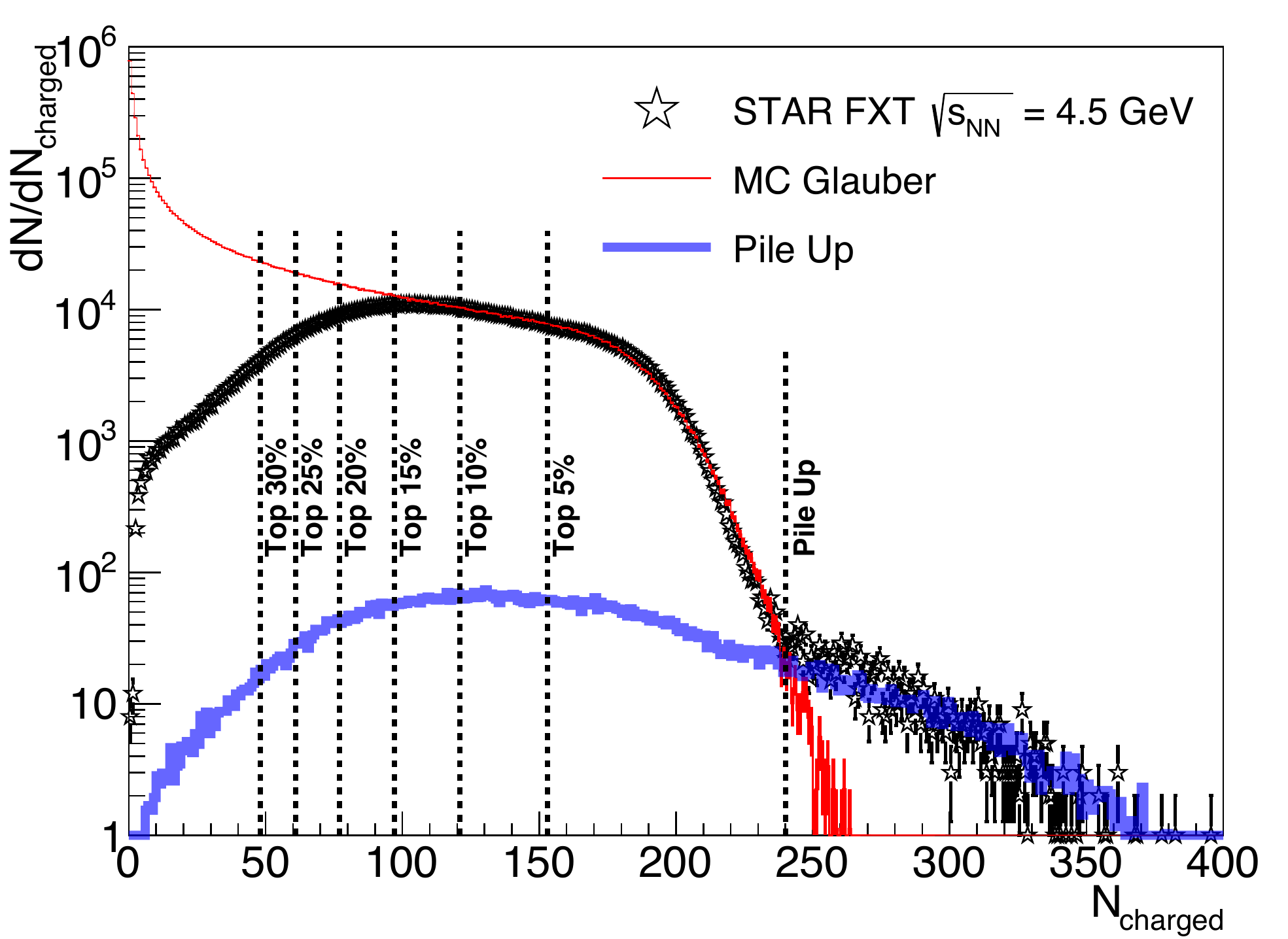}
\caption{\label{fig:centralityFigure} (Color online) Centrality selection for STAR FXT $\sqrt{s_{\mathrm{NN}}}$ = 4.5 GeV Au+Au collisions. The centrality variable $N_{\rm charged}$ is the number of tracks that pass the basic track cuts. The black points are the data, the thin red curve is the combined Monte-Carlo Glauber and negative binomial fit to the data, and the thick blue line is a Monte-Carlo model of pile-up events \cite{Meehan}. Vertical lines indicate the minimum number of tracks required for an event to be in the corresponding centrality bin. Events with multiplicity greater than 240 are dominated by pile-up, and are excluded from all analyses.}
\end{figure}

\begin{table}[ht]
\caption{The centrality selection used in the analyses. Included are the average number of participating nucleons ($N_{\rm part}$) estimated for the data for each centrality, the values of $N_{\rm part}$ predicted from a Glauber model for a minimum-bias trigger, the percentage of triggers corresponding to pile-up of two lower-multiplicity collisions, and the total number of events recorded.  Each centrality corresponds to 5\% of the total cross section. 
}
\label{tab:centrality_table}
\begin{tabular*}{0.5\textwidth}{@{}l*{15}{@{\extracolsep{0pt
          plus12pt}}l}}
Centrality & $\langle N_{\rm part}\rangle$ & $\langle N_{\rm part}\rangle$ & Pile-up & Events \\
(\% of $\sigma_{\rm total}$) & (Estimated) & (Min bias) & (\%) & \\
\hline
 0 -  5 & 341 $\pm$ 5 & 336 & 1.35 & 266,694 \\
 5 - 10 & 289 $\pm$ 9 & 286 & 0.72 & 267,347 \\
10 - 15 & 244 $\pm$ 8 & 242 & 0.58 & 258,854 \\
15 - 20 & 210 $\pm$ 6 & 204 & 0.49 & 203,600 \\
20 - 25 & 178 $\pm$ 5 & 170 & 0.44 & 125,539 \\
25 - 30 & 154 $\pm$ 4 & 142 & 0.40 &  68,844 \\
\hline
\end{tabular*}
\end{table}



The event selection cut requires the primary vertex to be within 1 cm of the target; 96.6\% of events pass this cut. Accepted tracks are required to have a distance of closest approach to the primary vertex of less than 3 cm (roughly six times the tracking resolution) and to include greater than half of the possible TPC hits to avoid double-counting of split tracks.

The distribution of charged particle multiplicities is shown in Fig.~\ref{fig:centralityFigure}.  Also shown in Fig.~\ref{fig:centralityFigure} are the centrality selection criteria. The centrality class and the average number of participating nucleons, labeled $\langle N_{\rm part}\rangle$ (minimum bias) in Table \ref{tab:centrality_table}, were estimated using a Monte Carlo Glauber model \cite{Ray_2008} assuming a negative binomial distribution for charged particle production. The Glauber model has been employed by STAR for centrality binning at collider energies from 200 to 7.7 GeV, and by the HADES collaboration for fixed-target Au+Au collisions at $\sqrt{s_{\rm NN}}$ = 2.4 GeV~\cite{Adamczewski-Musch:2017sdk}.
Comparison of the Glauber Monte Carlo and the data indicates that the trigger efficiency approaches unity for the most central collisions, and therefore we take this as an assumption and estimate the trigger efficiencies for less central collisions from the ratio of the number of recorded events over 267,000 (the average number of events for the two most central bins). For the 0-5\%, 5-10\%, 10-15\%, 15-20\%, 20-25\% and 25-30\% bins, the efficiencies are 100\%, 100\%, 97\%, 76\%, 47\% and 26\%, respectively. 
Overall, the trigger selects events corresponding to 22.5\% of the minimum-bias distribution. 
The estimated $\langle N_{\rm part}\rangle$ for each bin is then determined by taking a weighted average of $N_{\rm part}$, with weights equal to the number of recorded events for a given $N_{\rm charged}$, calculated as a function of $N_{\rm charged}$ from the Glauber model~\cite{doi:10.1146/annurev.nucl.57.090506.123020}. The uncertainty on the estimated $\langle N_{\rm part}\rangle$ values arises primarily from the central trigger which did not constrain the Glauber fits at low multiplicity.
Also shown in Fig.~\ref{fig:centralityFigure} is the estimated contribution of events which were the result of the pile-up of a triggered event along with a second minimum-bias collision in the target from the same bunch. Our estimate of the overall pile-up rate for all triggers is 0.8\%, which is consistent with 
there being a 20\% probability of having a gold ion incident on the target with each passing beam bunch. 
This pile-up probability is cross-checked and confirmed by measuring the number of vertices reconstructed from collisions one filled bunch after the triggered collision.
Due to the momentum resolution of the tracks and the projection distance back to the target (0.5 to 3.0 meters), the average distance of closest approach of a primary track to its vertex of origin is several millimeters. Thus, tracks from two separate collisions within the target would be reconstructed as emerging from a single vertex. 

\begin{figure}[th]
\includegraphics[width=0.30\textwidth]{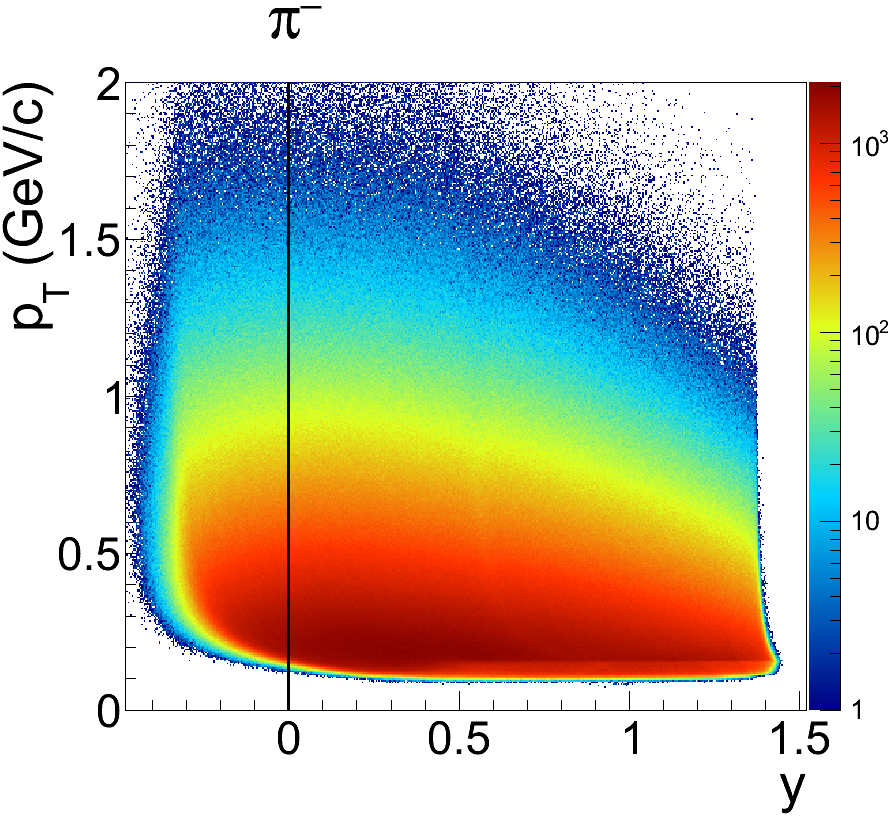}
\includegraphics[width=0.30\textwidth]{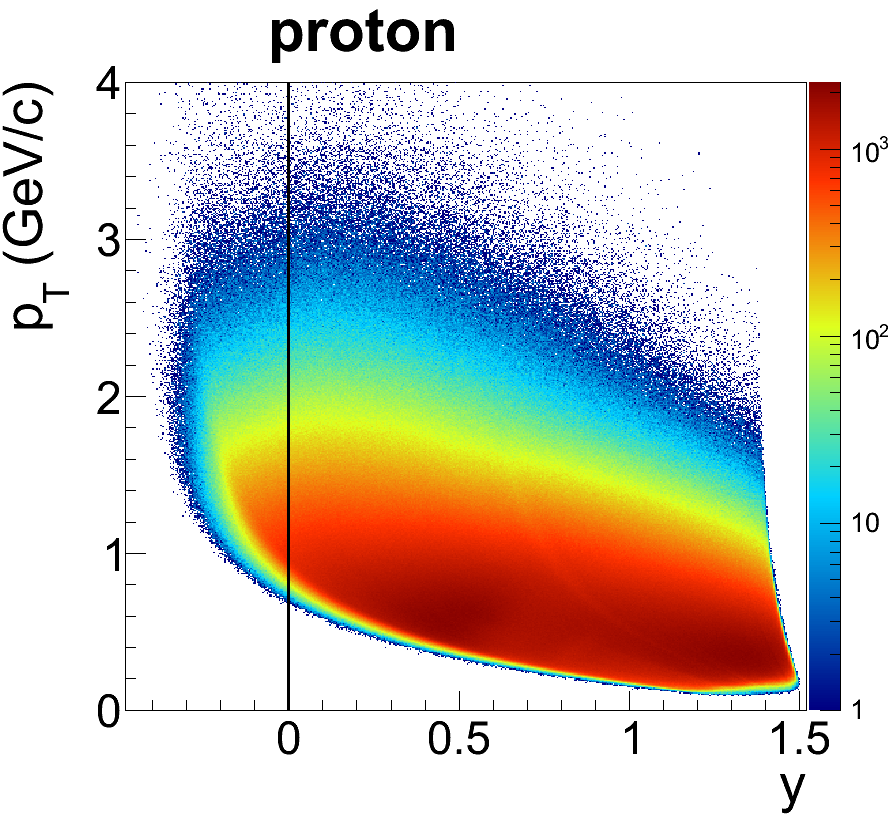}
\caption{\label{fig:QA_acceptance} Negative pion and proton relative yield versus rapidity and transverse momentum for STAR FXT $\sqrt{s_{\text{NN}}}$ = 4.5 GeV Au + Au collisions. The black line indicates the location of midrapidity.  The target (beam) rapidity in the center of mass
frame is at +1.52 (-1.52).}
\end{figure}

The location of the target along the beam axis was chosen to be $z = 211$ cm (where $z = 0$ corresponds to the center of the detector) in order to maximize the acceptance of the STAR Time Projection Chamber (TPC) \cite{ANDERSON2003659} for fixed-target events. Protons and pions were selected from all charged tracks within a 2$\sigma$ band centered on the Bichsel prediction for $dE/dx$ \cite{Bichsel:2006cs}. The acceptance effects are illustrated in Fig.~\ref{fig:QA_acceptance} by the distribution of the measured $p_T$ and rapidity, $y$, for protons and pions. For both the pions and protons, the right-hand edge is the $\eta_{\rm lab}$ = 0.1 acceptance limit, while the left-hand edge illustrates the $\eta_{\rm lab}$ = 2 acceptance limit. The magnetic field of the solenoid defines the low $p_T$ limit of 100 MeV/$c$. The detector does not impose a high $p_T$ limit; the high $p_T$ fall-off exhibited in Fig.~\ref{fig:QA_acceptance} is due the exponential production. For pions, there is good acceptance from midrapidity ($y = 0$) to beam rapidity ($y = 1.52$), while for protons, the $\eta_{\rm lab} = 2$ acceptance limit imposes a varying low $p_T$ limit. Geometric acceptances for charged kaons would fall between those of pions and protons, but, as seen in Fig.~\ref{fig:PID_map}, particle identification using $dE/dx$ would be limited to $p_{\rm total} < 600$ MeV/$c$, precluding analysis of midrapidity charged kaons. In this paper, the rapidity of a particle is always given in the collision center-of-momentum frame, not the laboratory frame.


 
\section{Directed Flow}

Characteristics of the QGP, including the nature of the transition between QGP and hadronic matter \cite{Steinheimer:2014pfa, Konchakovski:2014gda, Ivanov:2014ioa, Nara:2016hbg, Singha:2016mna, Nara:2017qcg, Nara:2020ztb}, can be explored via measurements of azimuthal anisotropy with respect to the collision reaction plane.  The reaction plane is defined by the beam axis and the vector connecting the centers of the two colliding nuclei. This anisotropy is characterized by a series of Fourier coefficients  \cite{Ollitrault:1992bk, Voloshin:1994mz, Poskanzer:1998yz, Bilandzic:2010jr}:
\begin{linenomath}
\begin{equation}
\label{eq:FourierCoeff}
v_n = \langle \cos n( \phi-\Psi_\text{R} ) \rangle,
\end{equation}
\end{linenomath}
where the angle brackets indicate an average over all events and particles of interest, $\phi$ denotes the azimuthal angle of each particle, $\Psi_\text{R}$ is the azimuthal angle of the reaction plane, and $n$ denotes the harmonic number. The sign of $v_1$ is positive for particles near the projectile rapidity, which is the same convention as used in fixed-target relativistic heavy-ion experiments at higher and lower beam energies. The present study explores the first two harmonics: directed flow ($v_1$) in the current section, and elliptic flow ($v_2$) in Section V. 

\subsection{Proton and pion $v_1$}

\begin{figure}[hbt]
\includegraphics[width=0.48\textwidth]{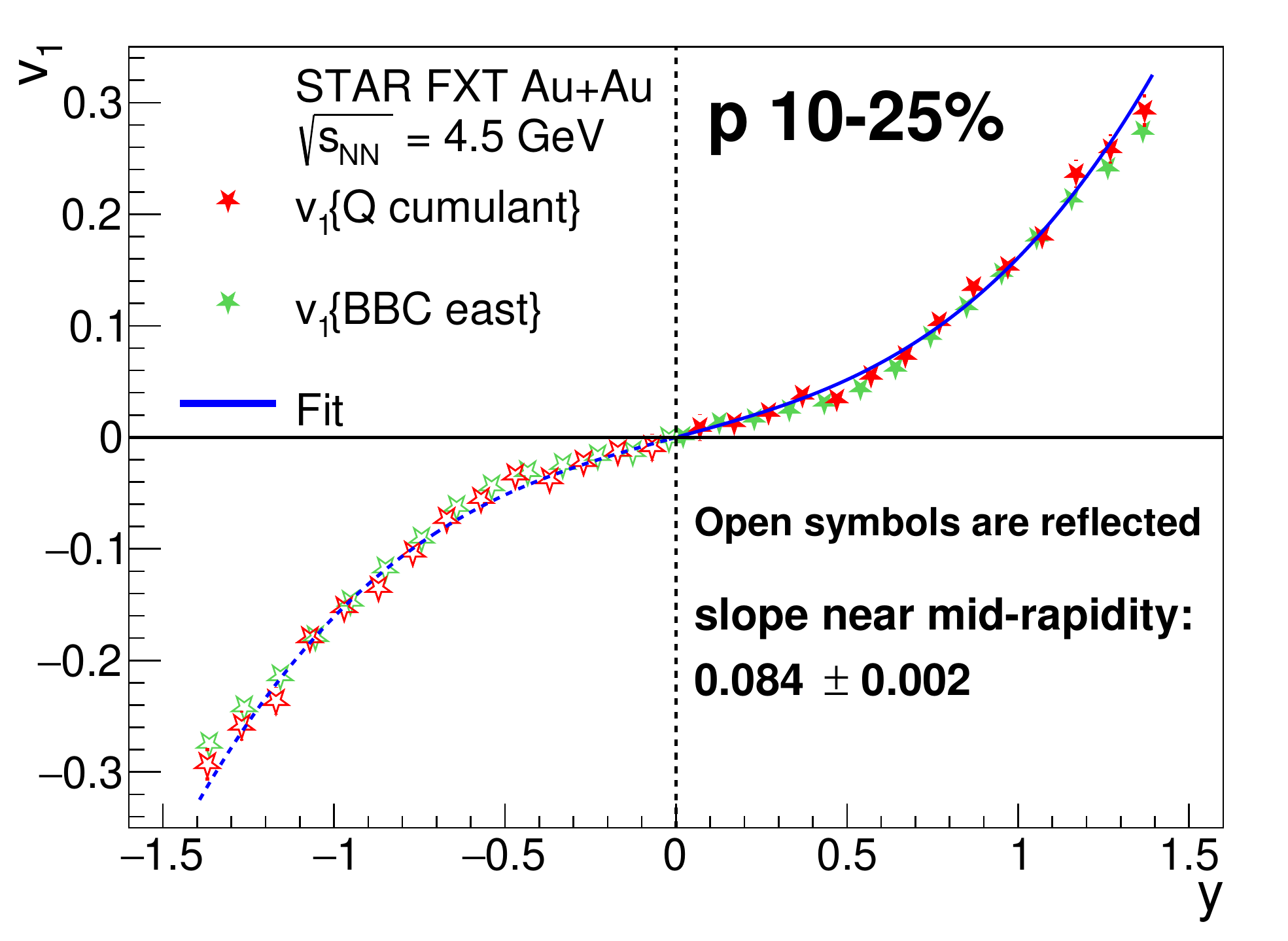}
\caption{
Rapidity dependence of directed flow, $v_1(y)$, for protons with transverse momentum $0.4 < \pt < 2.0$ GeV/$c$ from events with 10-25\% centrality. Two analysis methods, as discussed in the text, are compared. Plotted error bars are statistical only, and systematic errors are of comparable size. The curve is a cubic fit to the data.} 
\label{fig:V1Proton}
\end{figure}

All directed flow analyses in this paper pertain only to rapidity-odd $v_1(y)$, which is a measure of the collective sideward deflection of emitted particles.  The rapidity-even correlation $v_1^{\rm even}(y)$ \cite{Teaney:2010vd, Luzum:2010fb} is not related to the reaction plane in mass-symmetric collisions, and originates from initial-state event-by-event fluctuations.

We consider three distinct analysis methods: first, the TPC event plane (EP) approach with random sub-events for EP resolution correction \cite{Ollitrault:1992bk, Voloshin:1994mz, Poskanzer:1998yz}; second, a method based on the use of the Beam Beam Counter (BBC) detector for event plane determination \cite{whitten2008beam, Agakishiev:2011id, Adamczyk:2014ipa}; and third, a direct calculation of multi-particle cumulants (the Q-cumulant method) \cite{Bilandzic:2010jr}. Both the first and second methods use equation (\ref{eq:FourierCoeff}) to calculate the directed flow with the value of $\Psi_R$ and its resolution estimated from a sub-event calculation based on information from either the TPC or the BBC \cite{Poskanzer:1998yz}.   
The first method is less favored due in part to its susceptibility to bias from non-flow (correlations unrelated to the initial geometry of the collision) \cite{Bilandzic:2010jr}, but is investigated in the present proton directed flow study because that was the method used in 2000 by the E895 collaboration \cite{Liu:2000am}.  However, due to momentum conservation effects \cite{Borghini:2002mv}, this first method suffers from a relatively large departure from the $v_1(y)$ odd function behavior required by symmetry, and only the second and third methods are presented in Fig. \ref{fig:V1Proton}. 

More specifically, the red star markers in Fig.~\ref{fig:V1Proton} present proton $v_1(y)$ based on a 4th-order direct Q-cumulant calculation \cite{Bilandzic:2010jr}, which suppresses the contribution from non-flow.  The tracks included in the analysis have transverse momentum $0.4 < \pt < 2.0$ GeV/$\textit{c}$, which matches the selection used by E895 at $\sqrt{s_{\text{NN}}} =$ 4.3 GeV \cite{Liu:2000am} and by STAR in collider mode at $\sqrt{s_{\text{NN}}} =$ 7.7 - 200 GeV \cite{Adamczyk:2014ipa}.  Our centrality selection is 10-25\%, which is consistent with the centrality reported by the E895 collaboration \cite{Liu:2000am}. 
Due to the restricted acceptance and particle identification performance of the STAR detector in FXT mode (see Fig.~\ref{fig:QA_acceptance}), measurements are reported for only one side of midrapidity, and the odd-function behavior of directed flow is used to reflect points to the missing rapidity region.  

The east-west asymmetry of FXT mode requires us to rely on the east BBC detector for the event plane estimation.  Sub-event correlations between the east inner BBC (covering pseudorapidity 3.3 to 5) and the TPC \cite{Poskanzer:1998yz} are used to correct for event plane resolution.  The averaged east BBC event plane resolution for the slightly wider 10-30\% centrality bin used in the pion directed flow analysis is $41.4 \pm 0.4$\%.

\begin{figure}[hbt]
\includegraphics[width=0.48\textwidth]{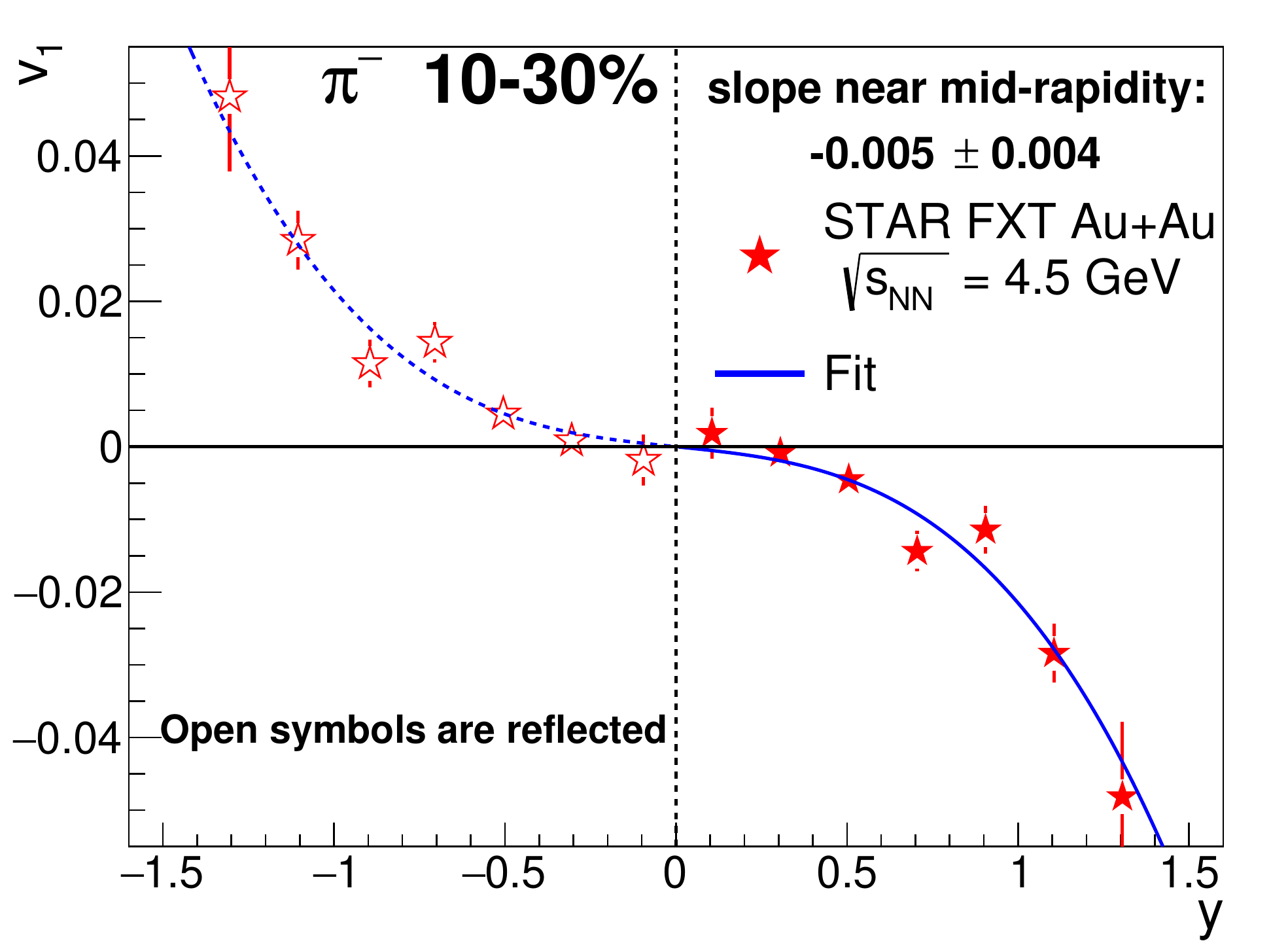}
\includegraphics[width=0.48\textwidth]{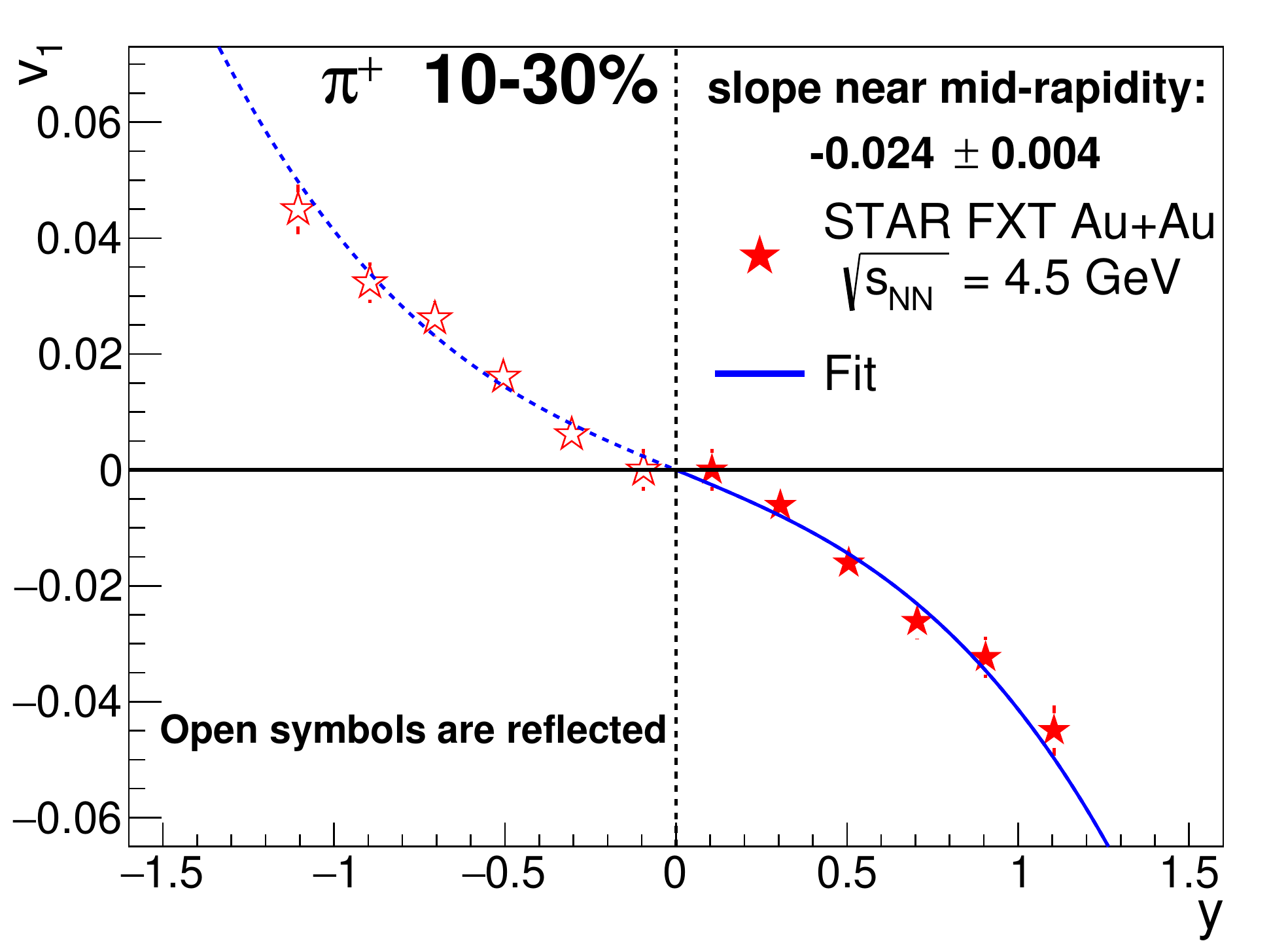}
\caption{
Upper panel: Rapidity dependence of directed flow, $v_1(y)$, for negative pions with transverse momentum $\pt > 0.2$ GeV/$c$ and total momentum magnitude $|p| < 1.6$ GeV/$c$ from events within 10-30\% centrality. Here, the BBC-based Event Plane method is used.  Plotted error bars are statistical only, and systematic errors are of comparable size. The solid curve is a cubic fit to the data. Lower panel: The same for positive pions. } 
\label{fig:V1Pion} 
\end{figure}

The shape of $v_1(y)$ at $\sqrt{s_{\text{NN}}} =$ 4.5 GeV is described quite well by a cubic function $F y + F_3 y^3$, where $F$ and $F_3$ are constants extracted from a fit to the data.  In order to study trends in proton directed flow as a function of beam energy, we take the linear term, $F = dv_1/dy \,|_{y = 0}$, to characterize the overall strength of the directed flow signal at each energy.  This is the same procedure as used at higher beam energies by STAR in collider mode \cite{Adamczyk:2014ipa} and at lower beam energies by E895 \cite{Liu:2000am}. 
The curve in Fig.~\ref{fig:V1Proton} shows the fit with $F$ and $F_3$ as free parameters. The extracted proton slope is $dv_1/dy \,|_{y = 0} = F = 0.084 \pm 0.002$
In Ref.~\cite{Adamczyk:2017nxg}, the directed flow slope for ten particle species is presented for Au+Au collisions at $\sqrt{s_{\text{NN}}} =$ 7.7 to 200 GeV. As some of the species in Ref.~\cite{Adamczyk:2017nxg} have relatively poor statistics, a more stable fit of the directed flow slopes in that analysis was obtained after requiring $F_3 = 0$.  For the purpose of a consistent comparison with the slopes reported in Ref.~\cite{Adamczyk:2017nxg}, we also report the extracted proton slope with $F_3 = 0$ in the present analysis, namely $F =  0.086 \pm 0.002$ based on a fit over $0 \leq y \leq 0.6$. 

Figure~\ref{fig:V1Pion} presents $v_1(y)$ for negative (upper panel) and positive (lower panel) pions using the BBC-based method referenced above.  The 4th-order direct Q-cumulant method, as employed in Fig.~\ref{fig:V1Proton}, provides consistent results, but in the context of the relatively poor statistics for charged pions in FXT mode at $\sqrt{s_{\text{NN}}} =$ 4.5 GeV, the statistical errors on the BBC-based method are significantly smaller.  No E895 $v_1$ measurements for pions were published, so the only available experimental data for comparison are STAR collider-mode measurements at $\sqrt{s_{\text{NN}}} =$ 7.7 GeV and above \cite{Adamczyk:2014ipa}.  While track selections of transverse momentum $\pt > 0.2$ GeV/$\textit{c}$ and total momentum magnitude $|p| < 1.6$ GeV/$\textit{c}$ match the measurements at higher energies, the limited centrality range of our 2015 FXT test run restricts the centrality in Fig.~\ref{fig:V1Pion} to 10-30\%, and does not fully match the 10-40\% centrality already published at $\sqrt{s_{\text{NN}}} =$ 7.7 GeV and above \cite{Adamczyk:2014ipa}.   
The blue line in Fig.~\ref{fig:V1Pion} shows the fit with $F$ and $F_3$ as free parameters. The extracted negative pion slope is $dv_1/dy \,|_{y = 0} = F = -0.005 \pm 0.004$ and positive pion slope is $dv_1/dy \,|_{y = 0} = F = -0.024 \pm 0.004$
For the purpose of a consistent comparison with slopes reported in Ref.~\cite{Adamczyk:2017nxg}, we also report the extracted negative and positive slopes with $F_3 = 0$ in the present analysis, namely $F = -0.013 \pm 0.003$ and $F = -0.032 \pm 0.003$, respectively, based on a fit over $0 \leq y \leq 0.8$.

The percentage difference between $\pi^+$ and $\pi^-$ directed flow becomes larger as we scan down from STAR collider energies to the present FXT energy point.  This observation is consistent with isospin or Coulomb dynamics becoming more prominent at lower beam energies, and is qualitatively consistent with measurements at even lower energies reported by the FOPI collaboration \cite{Reisdorf:2006ie}.  

Systematic errors arising from event-vertex cuts, particle ID cuts, and from contamination by other particle species, all make small to negligible contributions.  Systematic errors arising from a cut on global distance of closest approach to the collision vertex, from the minimum number of hits required for $dE/dx$ calculation, from the sensitivity to the fit range used when determining $dv_1/dy$, and from a correction for a region of diminishing proton acceptance near midrapidity, contribute at a level that is comparable to statistical errors.



\subsection{Lambda and Kaon $v_1$} 

Standard topological cuts on $\pi^+\pi^-$ and $p\pi^-$ pairs were utilized to identify $K^0_\text{S}$ mesons and $\Lambda$ baryons, 
respectively.
Events with 10-30\% centrality were selected for this 
analysis. The statistics of both $K^0_\text{S}$ and $\Lambda$ 
candidates are sufficient for the BBC or TPC event plane 
method with $\eta$-separated sub-events where the directed 
flow is calculated using Eq. (\ref{eq:FourierCoeff}).   
Two sub-event methods are used in this analysis.   
First, the event plane is reconstructed using BBC information (BBC event plane), and second, the event plane is reconstructed using primary protons and deuterons measured in the TPC with laboratory pseudorapidity $-0.9<\eta_\text{lab}<0$ for every $K^0_\text{S}$ or $\Lambda$ candidate (TPC event plane). In the TPC event plane method, protons originating from $\Lambda$ candidates are excluded from the event plane estimation in order to eliminate self-correlation between $\Lambda$ candidates and the event plane. Both TPC and BBC event plane resolutions  
are estimated using the method of three subevents \cite{Poskanzer:1998yz}. The TPC event plane resolution is estimated to be $67.5\pm0.5\%$ and the BBC event plane resolution to be $40.0\pm0.5\%$. 
The TPC event plane resolution can also be  calculated \cite{Poskanzer:1998yz} using the measured $v_1$ and multiplicity of protons and deuterons that are used to reconstruct the event plane. With an assumption that $v_1$ for deuterons is twice as large as for protons \cite{Wang:1994rua}, the calculated resolution is 70.2\%. 


\begin{figure}[th]
\includegraphics[width=0.48\textwidth]{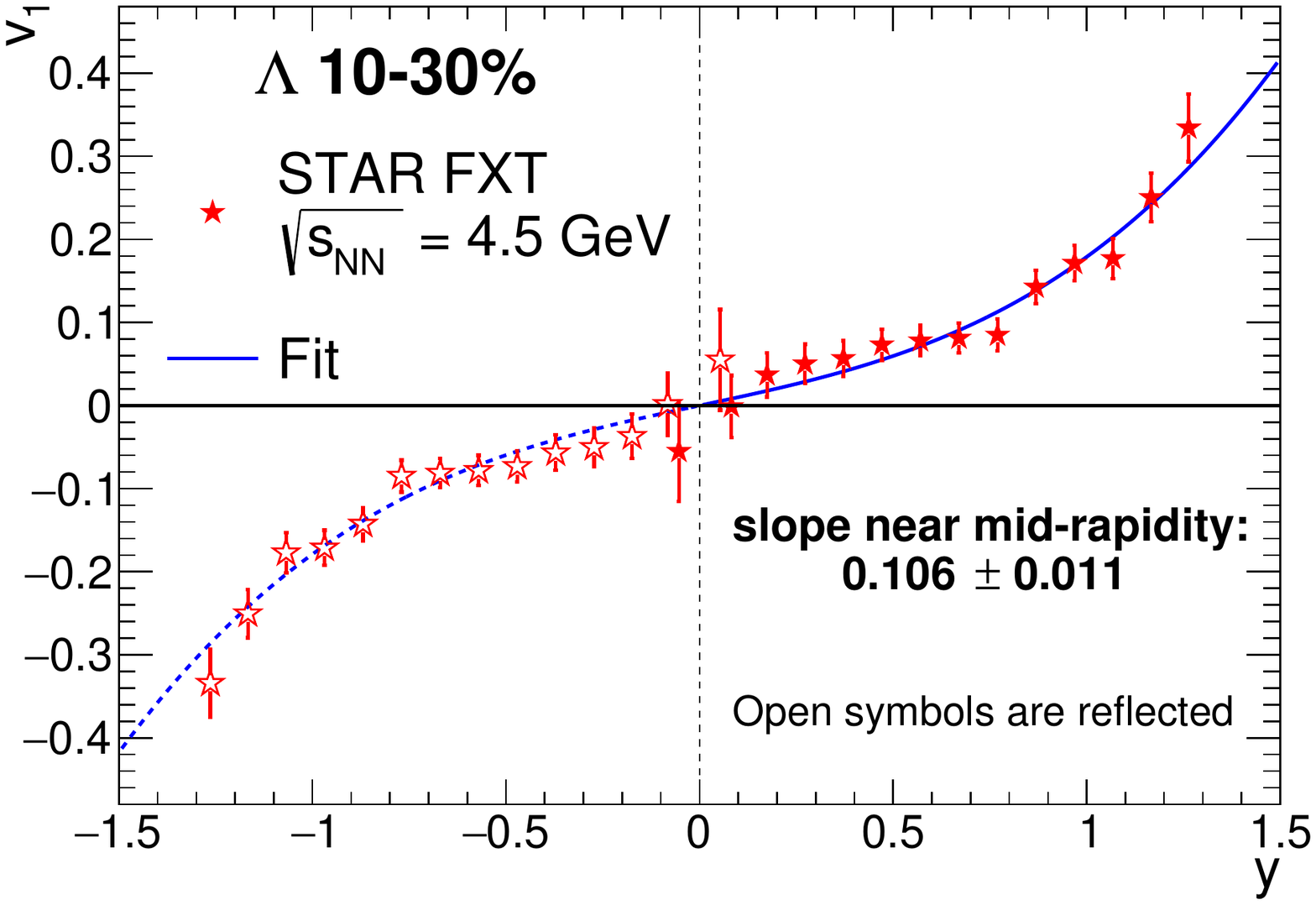}
\caption{
The rapidity dependence of the directed flow for the $\Lambda$ using the TPC event plane. Open symbols are the reflection of the solid symbols. The solid blue line is a cubic fit to the measured data. Plotted error bars are statistical only, while systematic errors are $\pm 0.7 \times 10^{-2}$.}
\label{fig:LambdaV1}
\label{fig:V1Lambda}
\end{figure}

\begin{figure}[th]
\includegraphics[width=0.48\textwidth]{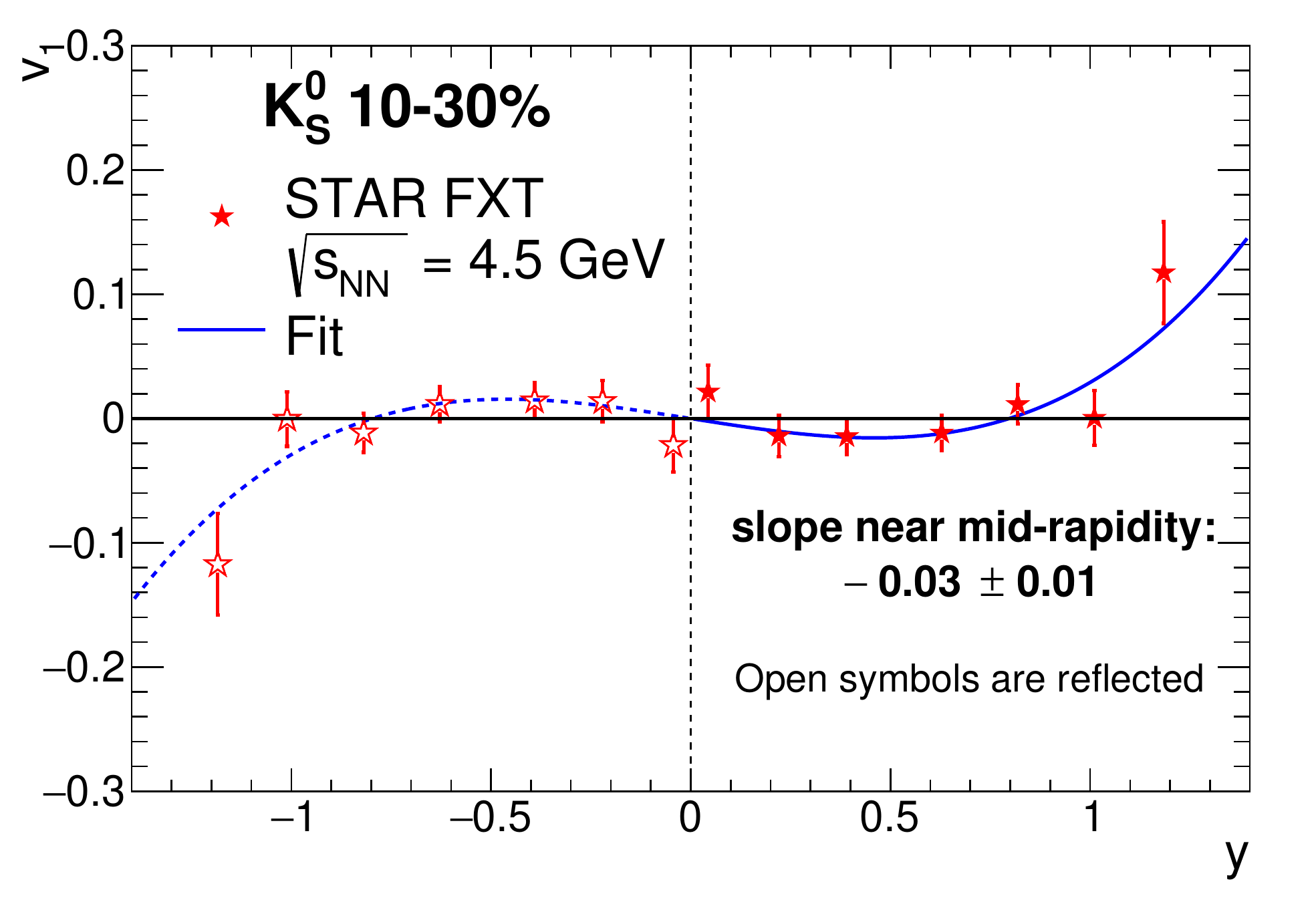}
\caption{
The rapidity dependence of the directed flow for the $K^0_\text{S}$ using the TPC event plane. Open symbols are the reflection of the solid symbols. The solid blue line is a cubic fit to the measured data. Plotted error bars are statistical only, while systematic errors are $\pm 1.7 \times 10^{-2}$.} 
\label{fig:V1K0s}
\end{figure}

The directed flow of $\Lambda$ or $K^0_\text{S}$ candidates is a superposition of a signal $v_1(y)$ and a background $v_1^B(y)$.  The combination is $v_1^{\rm tot}(y) = v_1(y)\Delta S + v_1^{B}(y)\Delta B$, where $\Delta S$ is the fraction (relative to the total) of the $\Lambda$ or $K^0_\text{S}$ signal and $\Delta B$ is the fraction of the combinatorial background accompanying the signal.
$\Delta S$ and its invariant mass resolution, $\sigma_\text{M}$, is calculated in every rapidity bin using the Pearson VII \cite{PearsonVIIdistribution} function fit to the invariant mass spectrum of either $\Lambda$ or $K^0_\text{S}$ candidates after the combinatorial background, whose yield is reconstructed using the momentum rotation technique \cite{Abelev:2010rv}, is subtracted. 
Using equation (\ref{eq:FourierCoeff}), the flow of the combinatorial background, $v_1^B(y)$, is calculated from particle pairs outside the mass region of the $K^0_\text{S}$ or $\Lambda$. 



Figure \ref{fig:V1Lambda} shows the directed flow of $\Lambda$ hyperons. The horizontal positions of the data points are corrected for the width of the bin. Six different sets of topological cuts are employed, varying the total number of $p\pi^-$ pairs from $\sim$540k to $\sim$160k, to observe how sensitive the directed flow of $\Lambda$ is to the size of the statistical sample. Two invariant mass windows $\pm2\sigma_M$ and $\pm 0.5\sigma_M$ are studied separately to vary the signal-to-background ratio, as well as the choice of either TPC or BBC event plane, to check if the event planes are consistent with each other. $v_1^B(y)$ is calculated in both cases in the $2<|\sigma_M|<5$
mass region outside of the center of the $\Lambda$ peak.
This gives a total of 24 results for slope parameters, $F$, representing the directed flow at midrapidity. 
Statistical errors on $v_1$ come from the upper and lower limit of slopes calculated using the covariance matrices of the cubic fits to the directed flow data. The weighted average from these 24 fits is $(10.6\pm1.1)\times 10^{-2}$ for $\Lambda$ hyperons. 
The systematic uncertainty, calculated as the average of the differences between the mean value of $10.6\times 10^{-2}$ and the nominal values from the fits, is $0.7\times 10^{-2}$.   

The directed flow of $K^0_\text{S}$ mesons was treated similarly, 
except wider 
binning was used and three invariant mass windows $\pm2\sigma_M, \pm 1\sigma_M$, and $\pm 0.5\sigma_M$. $v_1^B(y)$ is calculated in all three cases in the $2<|\sigma_M|<5$
mass region outside of the center of the $K^0_\text{S}$ peak.
In total, $\sim$110k $\pi^+\pi^-$ pairs pass the tightest topological cuts, while $\sim$370k pairs pass the loosest topological cuts. The weighted average of the total of 36 slope parameters $F$ is 
$(-3.4\pm1.1)\times 10^{-2}$ for $K^0_\text{S}$ and the systematic uncertainty is $1.7\times 10^{-2}$. The data points corrected for the bin widths are shown in Fig. \ref{fig:V1K0s}.

\subsection{Beam Energy Dependence}

\begin{figure}[th]
\includegraphics[width=0.48\textwidth]{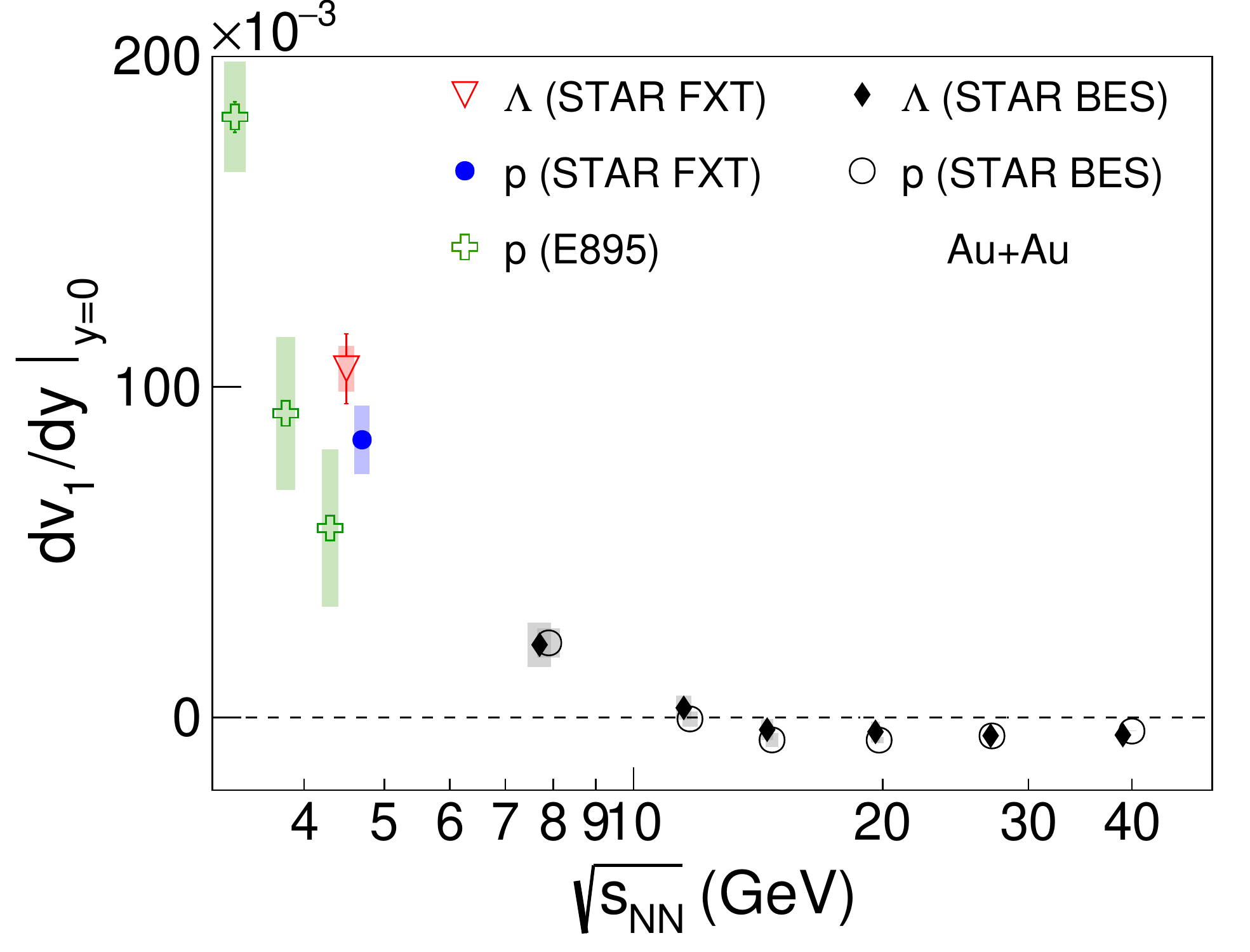} \\
\includegraphics[width=0.48\textwidth]{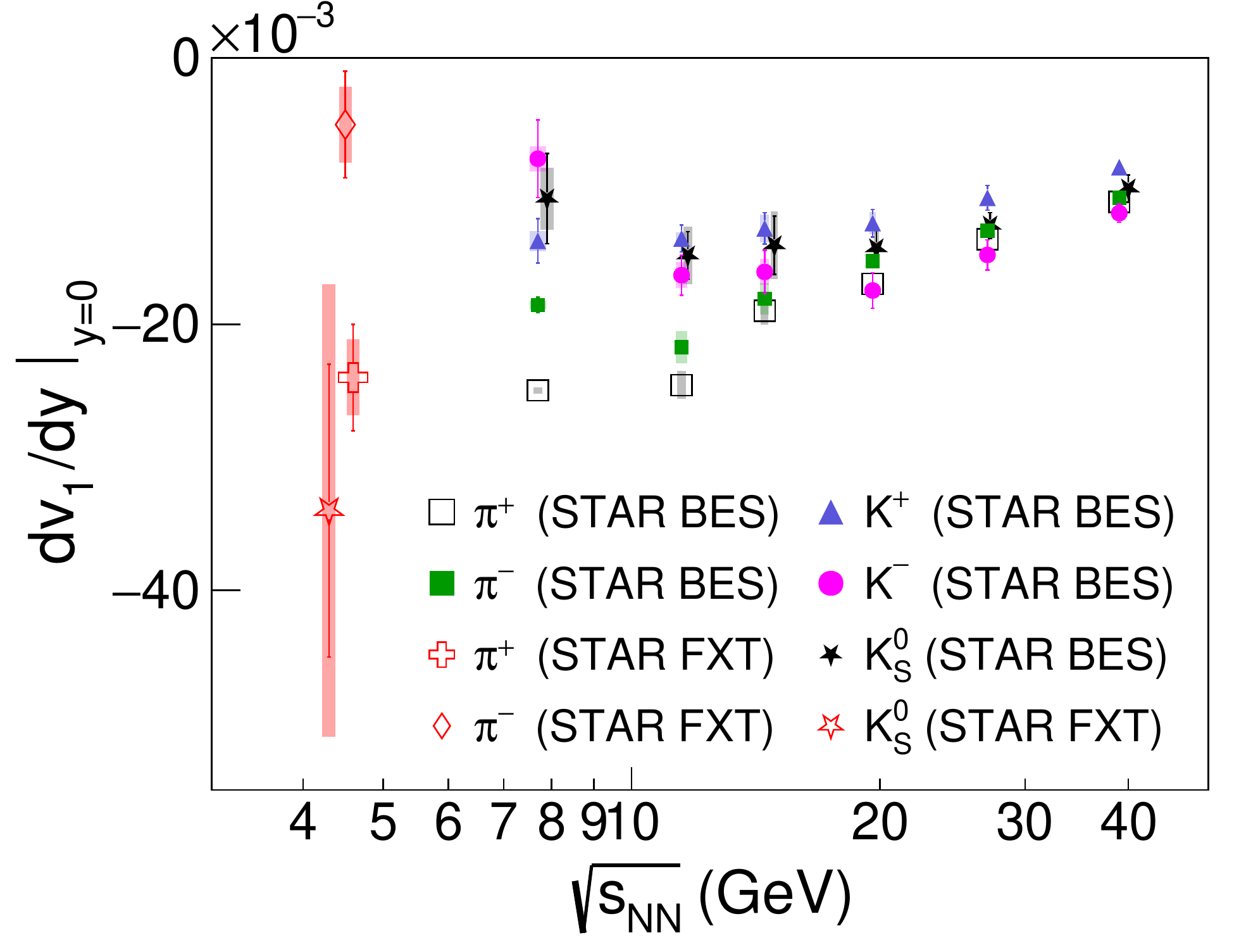}
\caption{(Color online) Beam energy dependence of the directed flow slope $dv_1/dy$ at midrapidity for baryons (upper plot) and mesons (lower plot) measured by STAR (this paper and Refs. \cite{Adamczyk:2014ipa,Adamczyk:2017nxg}) and by AGS experiment E895 \cite{Liu:2000am,Chung:2001je}. Some points are slightly offset horizontally.}
\label{fig:v1_slopes}
\end{figure}

Figure \ref{fig:v1_slopes} presents slopes $dv_1/dy \,|_{y = 0}$, based on the above-described cubic fits, for five species ($p$, $\Lambda$, $K^0_\text{S}$, $\pi^+$ and $\pi^-$) measured in Au+Au collisions in FXT mode at $\sqrt{s_{\text{NN}}} =$ 4.5 GeV.  Error bars show statistical uncertainties and shaded bands show systematic errors. The latter ones include factors already noted, as well as allowance for the rapidity range used in slope fitting.  

Liu {\it et al.} \cite{Liu:2000am} reported proton directed flow at centrality 12-25\% from the AGS E895 experiment, in the form of mean in-plane $p_T$ and $v_1(y)$ at $\sqrt{s_{NN}} = 4.3$ GeV and below.  In order to compare $dv_1/dy \,|_{y = 0}$ between STAR and E895, it is necessary to carry out a cubic fit to E895 $v_1(y)$ for protons using similar criteria as for STAR $v_1(y)$. The E895 fitted slopes in the upper plot of Fig. \ref{fig:v1_slopes} show statistical and systematic errors, where the latter arise from details of the fit. The E895 proton slopes reproduced in Ref. \cite{Adamczyk:2014ipa} are different, although consistent within errors, in part because Ref. \cite{Adamczyk:2014ipa} assumed errors on E895 $v_1(y)$ points that were equal to the marker size in cases where the actual errors were smaller than the published markers.   

Note that the new proton $v_1(y)$ slope measurement at $\sqrt{s_{\text{NN}}} =$ 4.5 GeV lies within errors on an interpolation between the same observable from STAR's published results for collider mode \cite{Adamczyk:2014ipa, Adamczyk:2017nxg} and E895 \cite{Liu:2000am}.  The highest E895 energy point at $\sqrt{s_{\text{NN}}} =$ 4.3 GeV agrees with the current FXT measurement within the uncertainties.  Proton and $\Lambda$ directed flow agree within errors at $\sqrt{s_{\text{NN}}} =$ 4.5 GeV. The $\Lambda$ directed flow results fit into a pattern that was observed by STAR at $\sqrt{s_{\text{NN}}} =$ 7.7 GeV and above \cite{Adamczyk:2017nxg}, but not at E895 energy points for $\sqrt{s_{\text{NN}}} =$ 3.8, 3.3 and 2.7 GeV \cite{Chung:2001je}.  

Positively charged pions, negative pions, and neutral kaons all show directed flow ($v_1$) signals in the opposite direction from that of the baryons, continuing trends observed at higher energies. The difference between $\pi^+$ and $\pi^-$ flow becomes stronger as the collision energy is reduced, which might be caused by isospin or Coulomb dynamics. 



\section{Elliptic Flow of Protons and Pions}
The second term in the Fourier decomposition of the azimuthal distribution, an elliptic flow $v_2$, of identified particles (protons and pions) measured in Au+Au collisions at $\sqrt{s_{\text{NN}}}$ = 4.5 GeV, is discussed in this section.  Elliptic flow of protons is compared with the earlier AGS data, while elliptic flow of pions has not been measured at this beam energy before. 
The appearance of number of constituent quark (NCQ) scaling, i.e.
the collapse of quark-number-scaled flow strengths for mesons and baryons onto a single curve, is considered to be evidence of QGP formation \cite{PhysRevLett.92.052302, Tian:2009wg}. Further and more detailed exploration of the energy region where NCQ scaling is not present is very interesting, as it might provide characterisation of relevant observables at the lower energies, where creation of QGP is in question. Protons, which have been analyzed at a similar energy by the E895 experiment at the AGS \cite{Pinkenburg:1999ya}, are compared to the previously published results from this experiment, while pions could only be compared to the results at higher energies. (Note that the results for protons at higher energies are published \cite{Adamczyk:2013gw, Adamczyk:2015fum}). Both positively and negatively charged pions are investigated separately in this analysis and it is found that they show the same behavior within uncertainties. Therefore, in the final plots positive and negative pions are presented together to improve the statistical significance of the result.   

In this analysis of elliptic flow, two methods are used:  (1) the event plane method using TPC information~\cite{Ollitrault:1992bk, Voloshin:1994mz, Poskanzer:1998yz} and (2) the two-particle cumulants method~\cite{Bilandzic:2010jr}. The event plane resolution is about 20$\%$. Resonance decays generate unrelated correlations of particles in the final state. Such correlations are a non-flow contribution and they bias the elliptic flow measurement. 
Since particles from resonance decays are correlated both in $\eta$ and $\phi$, we can reduce the non-flow contribution caused by resonances by measuring elliptic flow using particles which are not correlated in $\eta$. The implementation of this idea is different in each method. For the event plane method, we divide each event into two sub-events.
For the cumulant method, we require a 0.1 gap in $\eta$ between all considered pairs. Both methods give results which are consistent within their uncertainties. 

\begin{figure}[th]
\includegraphics[width=0.48\textwidth]{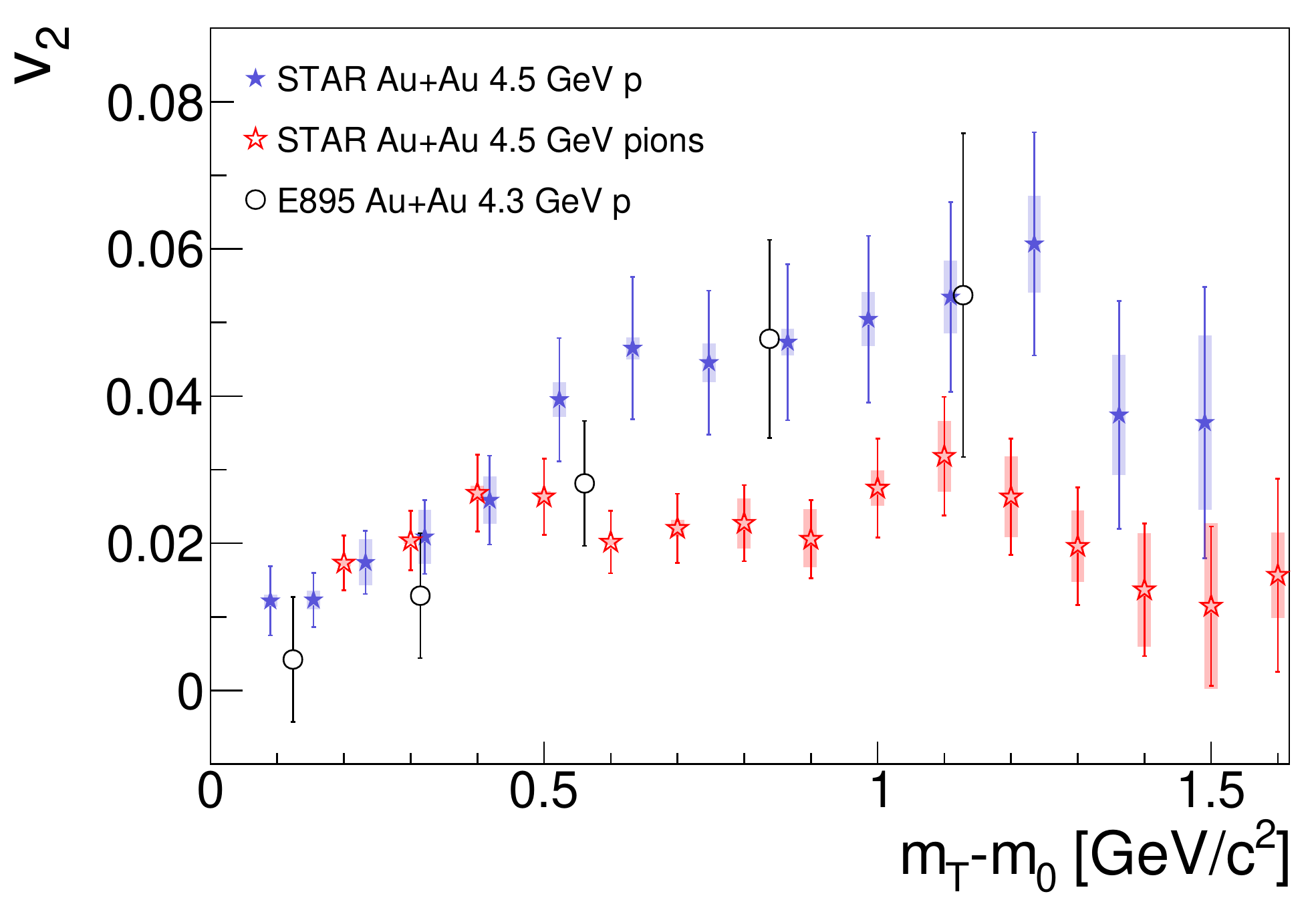}
\caption{$v_2$ of protons and pions from STAR FXT data analysis, and $v_2$ of protons from E895 experiment. Blue (red) stars represent STAR FXT proton (pion) data (0-30\% centrality), and black circles show E895 data (12-25\% centrality)~\cite{Pinkenburg:1999ya}.}
\label{fig:V2_FTX_to_E895}
\end{figure}

Figure~\ref{fig:V2_FTX_to_E895} shows the elliptic flow $v_2$ as a function of transverse kinetic energy $m_\text{T} - m$ for pions and protons obtained with the event plane method, where $m$ is mass and $\mt = \sqrt{m^2 + p_\text{T}^2}$ is transverse mass. It is compared to E895 results~\cite{Pinkenburg:1999ya} obtained using the same method. We analyze the 0-30\% most central events. For pions and protons, we require $|y| < 0.5$. In this analysis, we use tracks with $0.2<\pt<2.0$~GeV/c,  but due to STAR acceptance in FXT mode at $\sqrt{s_{\text{NN}}}$ = 4.5 GeV, we could analyze only protons with higher values of $\pt$, namely $\pt>0.4$~ GeV/c (see Fig.~\ref{fig:QA_acceptance}). The proton results are consistent with E895 results~\cite{Pinkenburg:1999ya}. 

\begin{figure}[th]
\includegraphics[width=0.48\textwidth]{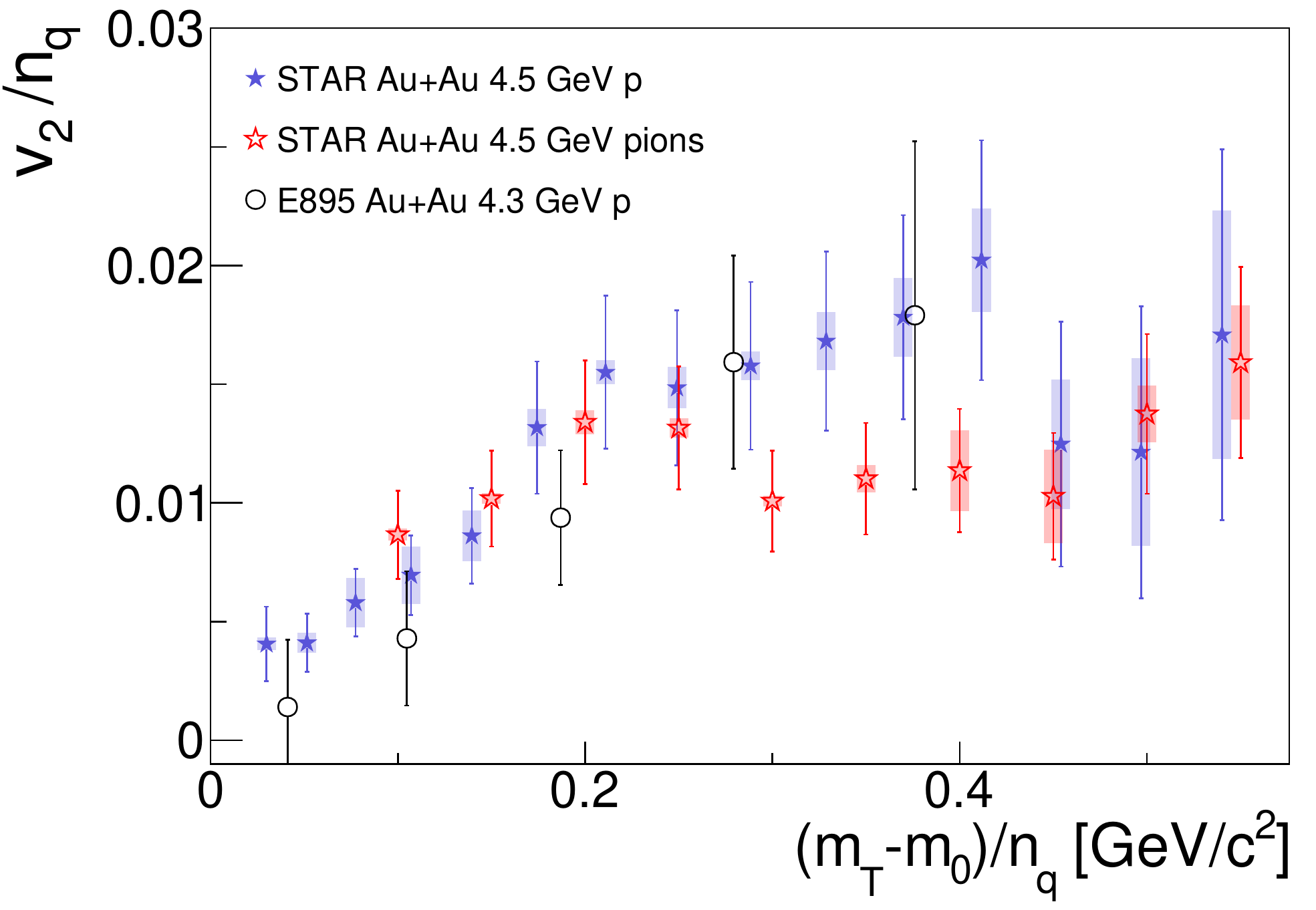}
\caption{$v_2$ scaled by the number of constituent quarks ($n_q$) for charged pions (red stars) and protons (blue stars) for 0-30\% central collisions. The values of $v_2$ scaled with $n_q$ for pions and protons are consistent with each other within errors. 
For comparison, points from E895 are also shown (black circles)}
\label{fig:V2ncq}
\end{figure}

To test the NCQ scaling, we divide $v_2$ and $m_\text{T} - m$ (Fig.~\ref{fig:V2_FTX_to_E895}) by the number of constituent quarks (3 for protons and 2 for pions). The results are presented in Fig.~\ref{fig:V2ncq}.   
The observed scaling with the number of constituent quarks at 4.5 GeV is similar to what is observed for Au+Au at higher collision energies~\cite{Adamczyk:2013gw,Adamczyk:2015fum}. The system created for Au+Au at $\sqrt{s_{\text{NN}}}$ = 4.5 GeV has, perhaps surprisingly, larger collectivity than expected, and there is no significant difference in identified particle elliptic flow behavior when compared to higher energies. 
The results in Figure 13 are in possible conflict with expectations.
Constituent-quark scaling 
  ($\tfrac{1}{3}v_2^{p}\left(m_T/3\right)=\tfrac{1}{2}v_2^{\pi}\left(m_T/2\right)$
  at intermediate $m_T$) at these energies would suggest partonic collectivity -- quark gluon plasma creation -- in Au+Au collisions at energies as low as $\sqrt{s_{NN}}=4.5$~GeV.
Higher statistical precision is needed to test the NCQ scaling hypothesis decisively, and this is forthcoming in the second phase of the beam energy scan.

\begin{figure}[th]
\includegraphics[width=0.48\textwidth]{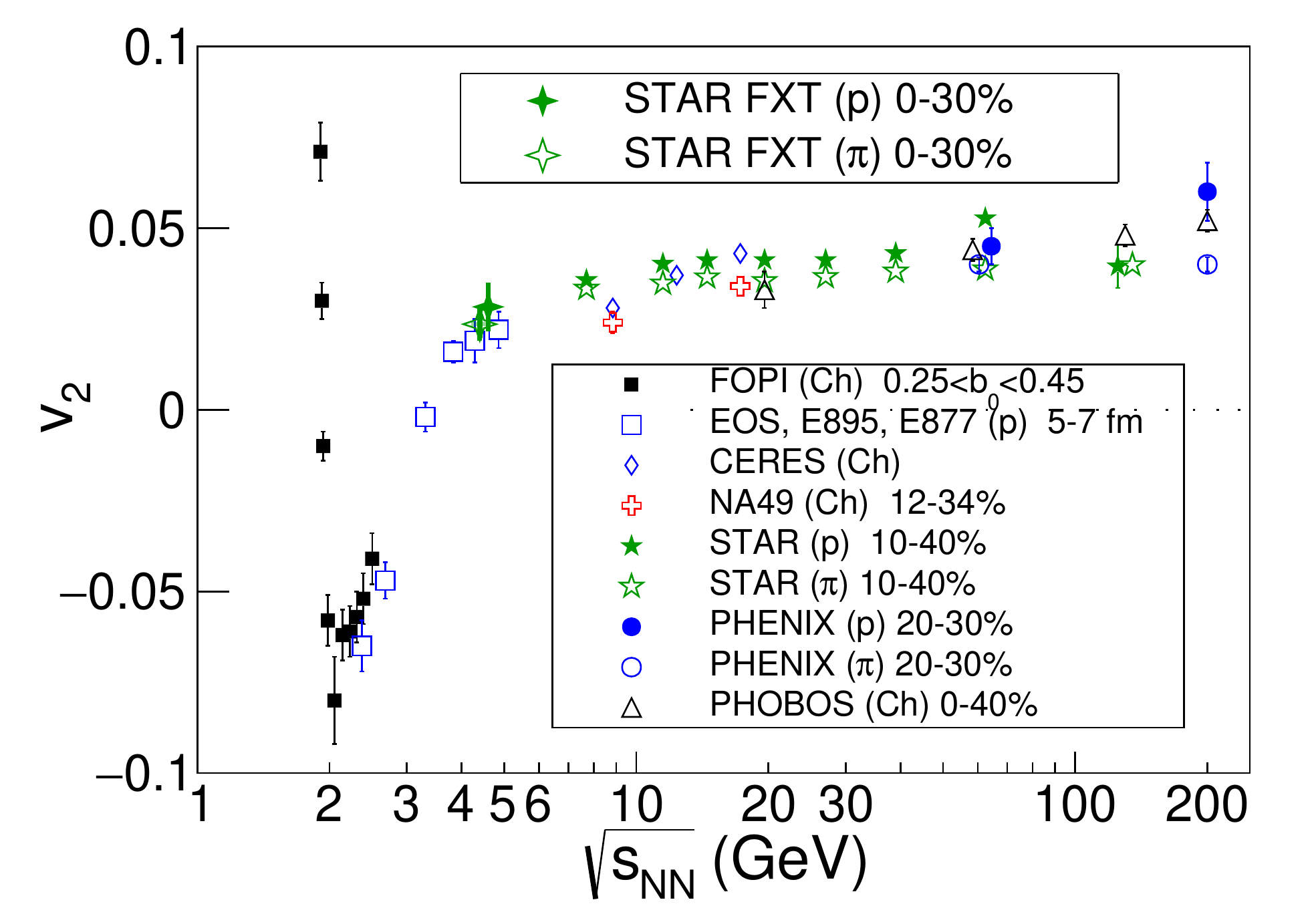}
\caption{The excitation function $v_2$ for all charged particles or separately for protons and pions, measured by several experiments. The STAR FXT points for protons and for pions are near the region where a change in slope occurs. 
Data are shown from FOPI~\cite{FOPI:2011aa,Andronic:2004cp}, 
E895~\cite{Pinkenburg:1999ya}, E877~\cite{BraunMunzinger:1998cg}, CERES~\cite{Adamova:2002qx}, NA49~\cite{Alt:2003ab}, PHENIX~\cite{Adare:2014bga}, PHOBOS~\cite{Back:2004zg}, and from the STAR collider energies~\cite{Adamczyk:2013gw,Adamczyk:2015fum,Adler:2001nb,Adams:2004bi,Adams:2004yc}. }
\label{fig:V2_systematics}
\end{figure}

Figure \ref{fig:V2_systematics} shows the beam energy dependence of $v_2$ measurements, integrated over $p_\text{T}$. The current results are consistent with the trends established by the previously published data.

\newcommand{\vecArrow}[1]{\overset{\rightarrow}{#1}}

\section{Femtoscopy of Pions}
\label{sec:HBTofPions}

Two-particle correlations at low relative momentum can be used to extract
information on the space-time structure of the particle-emitting
source.
Femtoscopy-- the technique of constructing and analyzing these correlations-- has been performed in heavy-ion experiments over a broad range of 
energies~\cite{Lisa:2005dd}.
In addition to providing a stringent test of the 
space-time structure of the final-state emission distribution
predicted
  by specific dynamical models~\cite{Lisa:2005dd}, the energy dependence of femtoscopic scales
  may reveal fundamental insights into the QGP equation of state.
As we discuss below, the low-energy results presented here help reveal
  a structure predicted~\cite{RISCHKE1996479,PhysRevD.33.1314} to probe
  the latent heat of the deconfinement transition.

\subsection{Methodology}
\label{sec:HBTMethodology}
Femtoscopic correlation functions are formed by making distributions of the relative momenta $ \vecArrow{q} \equiv \vecArrow{p_{1}} - \vecArrow{p_{2}}$ of pairs of particles.
A numerator distribution $N(\vecArrow{q})$ is formed using pairs where both tracks are from the same event, while a denominator distribution, $D(\vecArrow{q})$, is formed by constructing pairs where the two tracks are from separate events, but having similar multiplicity and positions of the primary vertex; this is known as the "mixed-event" technique~\cite{Kopylov:1972qw, Heinz:1999rw}.
The shape of both distributions will be dominated by the two-particle phase space distribution, but $N(\vecArrow{q})$ will also contain contributions from Coulomb interactions and Bose-Einstein effects.
The correlation function is the ratio

\begin{equation}
\label{eq:CFExpDefinition}
    C(\vecArrow{q}) = \frac{N(\vecArrow{q})}{D(\vecArrow{q})} .
\end{equation}
This ratio is sensitive to the space-time structure of the pion emitting source \cite{Goldhaber:1960sf,Lisa:2005dd}.

\begin{figure}[th]
\includegraphics[width=0.48\textwidth]{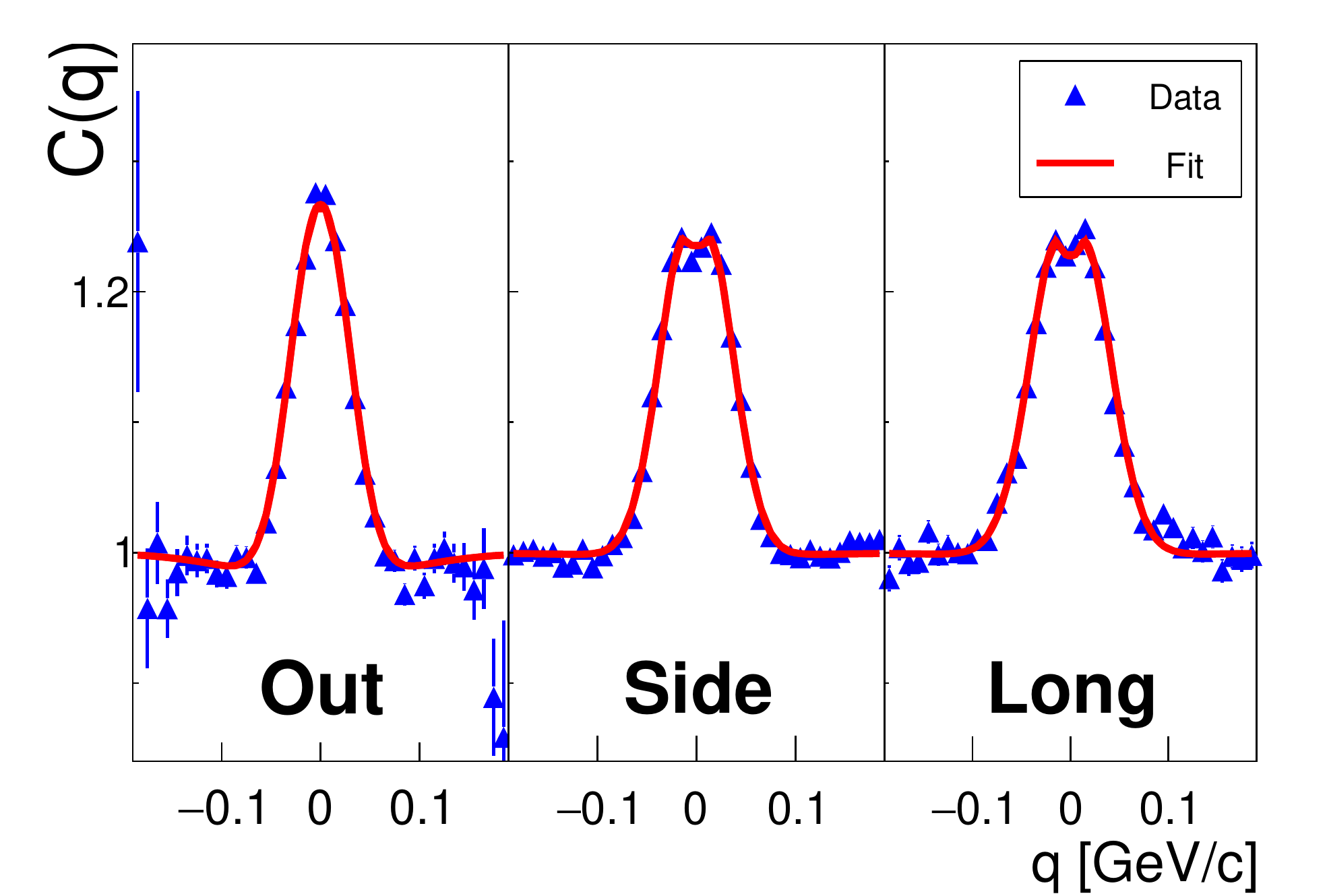}
\caption{
    Projections of the correlation functions in the LCMS frame onto the $q_\text{out}$, $q_\text{side}$, and $q_\text{long}$ axes for $\pi^-\pi^-$ pairs from events in the 0-10\% centrality range.
    Pairs are created from tracks in the momentum range 0.1 $ < \pt < $ 0.3 GeV/$\textit{c}$.
    For each projection $q_i$ shown, the other components of relative momentum
    are integrated over the range $|q_j|<35$~MeV/c.
    The red curve shows the projections~\cite{Lisa:2005dd} of a 3-dimensional fit to 
    equation~\ref{eq:CFTheoryDefinition}.  
    Errors are statistical only.
}
\label{fig:fig19}
\end{figure}

\begin{figure}[th]
\includegraphics[width=0.48\textwidth]{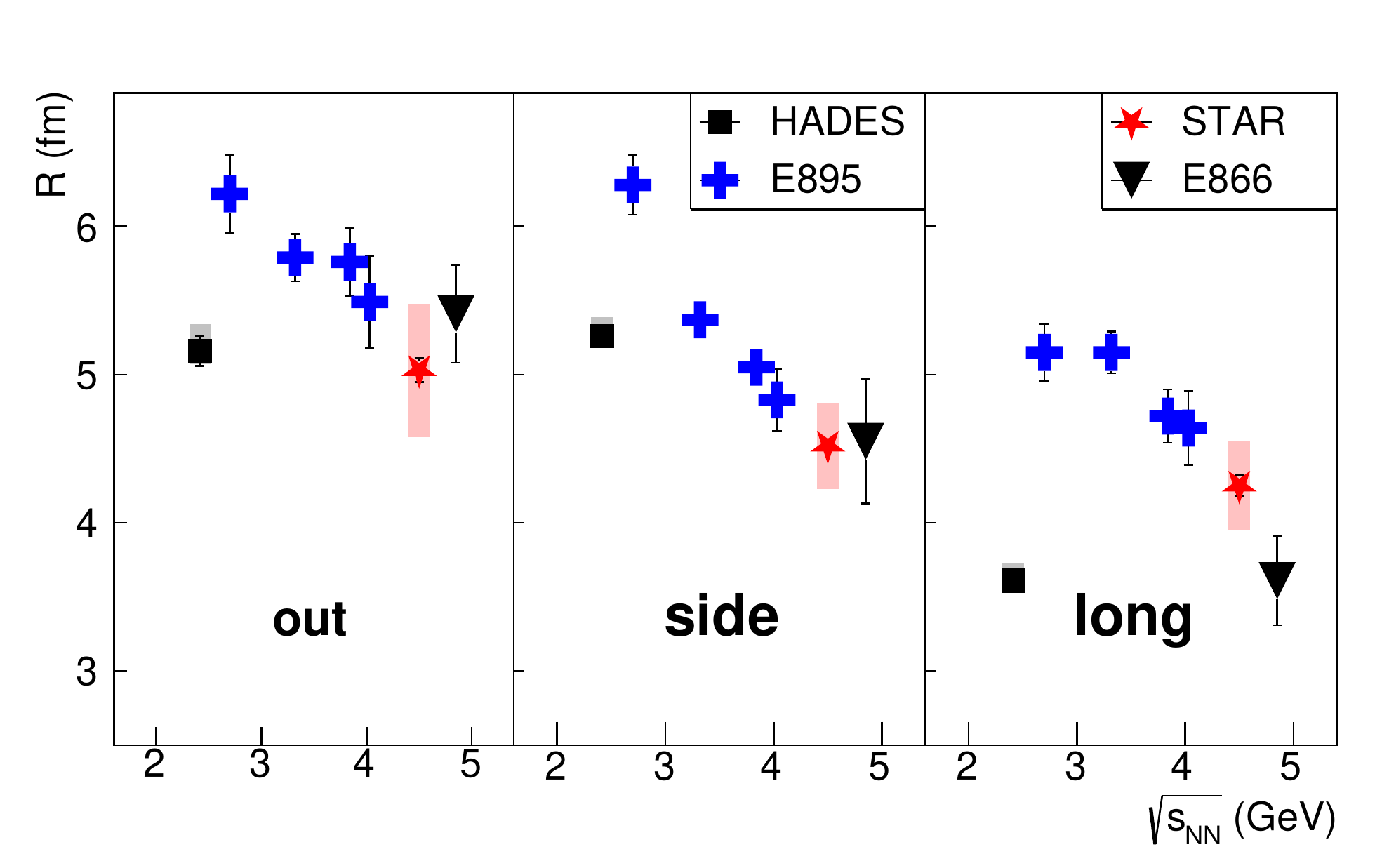}
\caption{
    Excitation function of \Rout, \Rside, and \Rlong for four experiments: HADES~\cite{Adamczewski-Musch:2018owj}, E895~\cite{PhysRevLett.84.2798}, STAR, and E866~\cite{Ahle:2002mi}.
    STAR points show both systematic (red boxes) and statistical errors (black lines) while errors for E895 and E866 are statistical only.  HADES systematic errors are roughly the same size as the datapoints.
    The same momentum and centrality selections are applied as in Fig. \ref{fig:fig19}.
}
\label{fig:fig20}
\end{figure}

Care must be taken to account for the effects of track reconstruction inefficiencies on the correlation function.
Single-track inefficiencies are common to both $N(\vecArrow{q})$ and $D(\vecArrow{q})$ and cancel in the ratio $C(\vecArrow{q})$.
However, two-track artifacts will affect $N(\vecArrow{q})$ alone, distorting $C(\vecArrow{q})$ at low $|\vecArrow{q}|$.
Track splitting (where hits from one charged particle are reconstructed as two distinct tracks) artificially enhances same-event pairs at low $ q $.
To eliminate this effect, we required both tracks to register separate hits on a minimum number
of pad rows~\cite{Adler:2001zd, Adams:2003ra, Adams:2004yc}.

Track merging (where hits from two charged particles are reconstructed as one track) suppresses same-event low-$q$ pairs.
These pairs cannot be recovered in the numerator $N(\vecArrow{q})$, but similar pairs can be removed from the mixed-event distribution $D(\vecArrow{q})$ to compensate.
To this end, we require all pairs to have a fraction of merged hits $ f_{\text{MH}} <$ 10\% \cite{Adler:2001zd, Adams:2003ra, Adams:2004yc}.
All pair cuts are applied equally to $N(\vecArrow{q})$ and $D(\vecArrow{q})$.

The relative momentum is evaluated in the Longitudinally Co-Moving System (LCMS), which is chosen such that $ (\vecArrow{p_{1}} +\vecArrow{p_{2}}) \cdot \hat{z} = 0 $, where $\hat{z}$ is the beam direction.
The relative momentum $ \vecArrow{q} $ is expressed in the Bertsch-Pratt \cite{PhysRevD.33.1314, PhysRevC.37.1896, PhysRevLett.74.4400} out-side-long coordinate system.
The ``longitudinal'' direction, $ q_{\text{long}} $, is taken to be the beam direction.
The ``out" direction, $ q_{\text{out}} $, is taken to be the direction of the transverse component of the pair-momentum $ \vecArrow{k}_{T} = (\vecArrow{p_{1}} + \vecArrow{p_{2}}) / 2 $, and the ``side", $q_{\text{side}}$, direction is defined to be perpendicular to the other two directions.

We use a Gaussian parameterization of the correlation function \cite{PhysRevC.66.044903} to relate the experimental quantity in eq. (\ref{eq:CFExpDefinition}) to the shape of the pion emitting source.
The correlation function that would arise solely from quantum statistical effects is represented by the quantity $C_{\text{free}}$ and can be expressed as
\begin{linenomath}
\begin{align}
    C_{\text{free}}(\vecArrow{q}) = 1 + &\exp\left(-R^{2}_\text{out}q^2_\text{out} - R^{2}_\text{side}q^2_\text{side} \right.\nonumber\\
    &  \left.- R^{2}_\text{long}q^2_\text{long}- 2R^2_\text{out-long}q_\text{out}q_\text{long} \right) .
\end{align}
\end{linenomath}
Here \Rout, \Rside, and \Rlong give the lengths of the regions of homogeneity~\cite{Makhlin1988} in the out, side, and long directions, respectively.
The cross term $ R^2_\text{out-long} $ represents a tilt of the correlation function in the $q_\text{out}-q_\text{long}$ plane.
To account for Coulomb interactions and contributions from halo pions we fit the data with the Bowler-Sinyukov functional form~\cite{BOWLER199169, SINYUKOV1998248, Adams:2004yc}:

\begin{equation} 
\label{eq:CFTheoryDefinition}
C(\vecArrow{q}) = (1-\lambda) + \lambda K(q_{\text{inv}})C_{\text{free}}(\vecArrow{q}) ,
\end{equation}
where $\lambda$ is the fraction of pion pairs that carry a correlation signal (as opposed to, for instance, non-primary pions from resonance decays which are uncorrelated with pions from the fireball, at the
resolution of our measurement).
Electromagnetic final state interactions are quantified by $K$, the spatially-integrated 
squared Coulomb wave function.
This function depends on the Lorentz invariant $q_{\rm inv}\equiv\sqrt{-q_{\mu}q^{\mu}}$,  where
$q_{\mu}=\left(E_1-E_2,\vec{q}\right)$.
The integral is taken over a spherical source 5~fm in radius \cite{Adams:2004yc, Abelev:2009tp}.
Integrating instead over a 3-fm source leads to negligible systematic error.

\begin{figure}[th]
\includegraphics[width=0.48\textwidth]{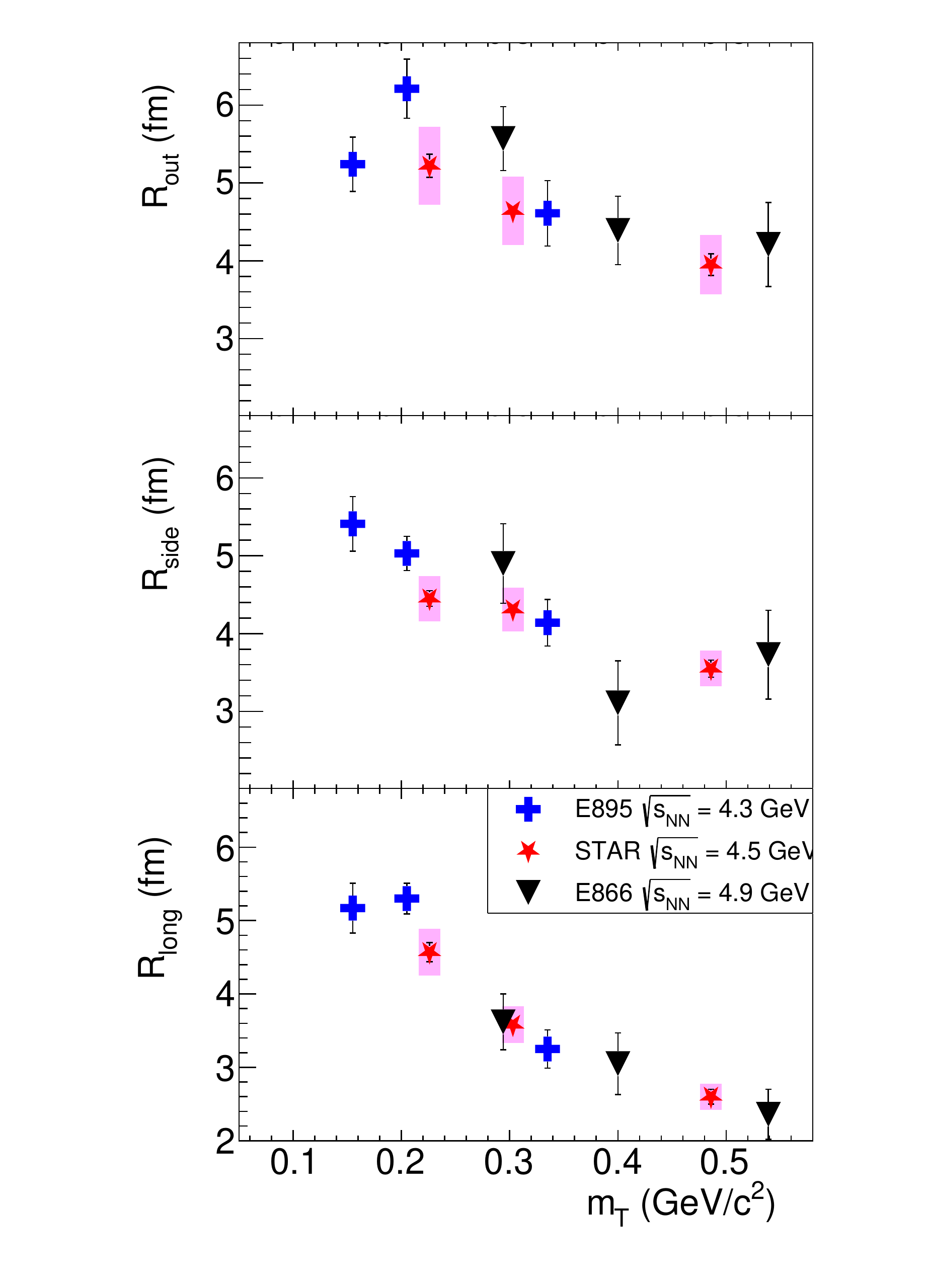}
\caption{
    Transverse mass dependence of \Rout, \Rside, and \Rlong for three experiments: E895~\cite{PhysRevLett.84.2798}, STAR, and E866~\cite{Ahle:2002mi}.
    Pairs for the STAR points are created from negative pion tracks in the momentum range 0.15 $ < \pt < $ 0.8 GeV/$\textit{c}$ from events in the 0-15\% centrality range. STAR points show both systematic (magenta boxes) and statistical errors (black lines) while errors for E895 and E866 are statistical only.
}
\label{fig:fig21}
\end{figure}


\subsection{Results}
\label{sec:HBTResults}

\begin{figure}[th]
\includegraphics[width=0.48\textwidth]{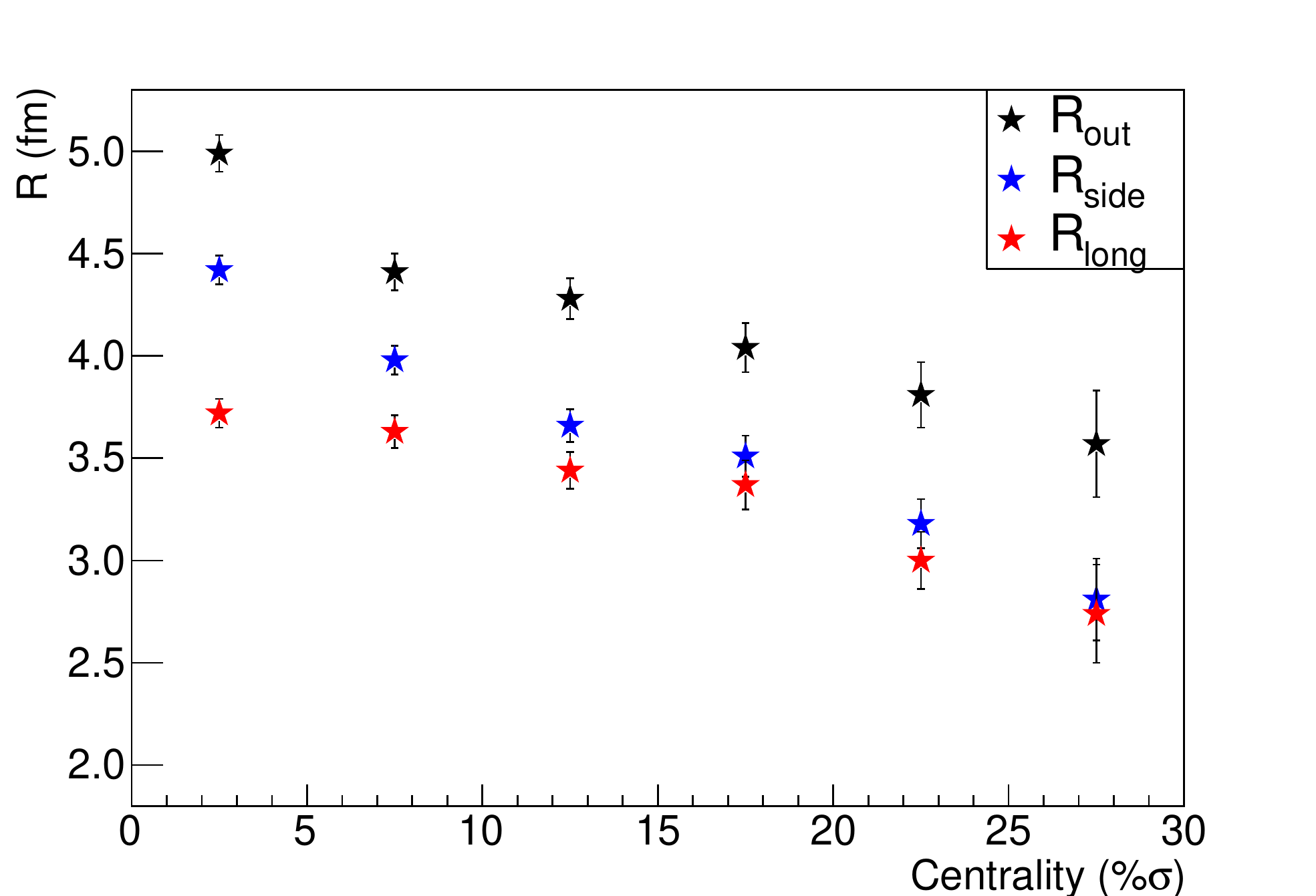}
\caption{
    The centrality dependence of \Rout, \Rside, and \Rlong.
    Errors are statistical only.
    Here $\pi^+\pi^+$ and $\pi^-\pi^-$ pairs in the momentum range 0.15 $ < \pt < $ 0.8 GeV/$\textit{c}$ are used.
}
\label{fig:fig22}
\end{figure}

\begin{figure}[th]
\includegraphics[width=0.48\textwidth]{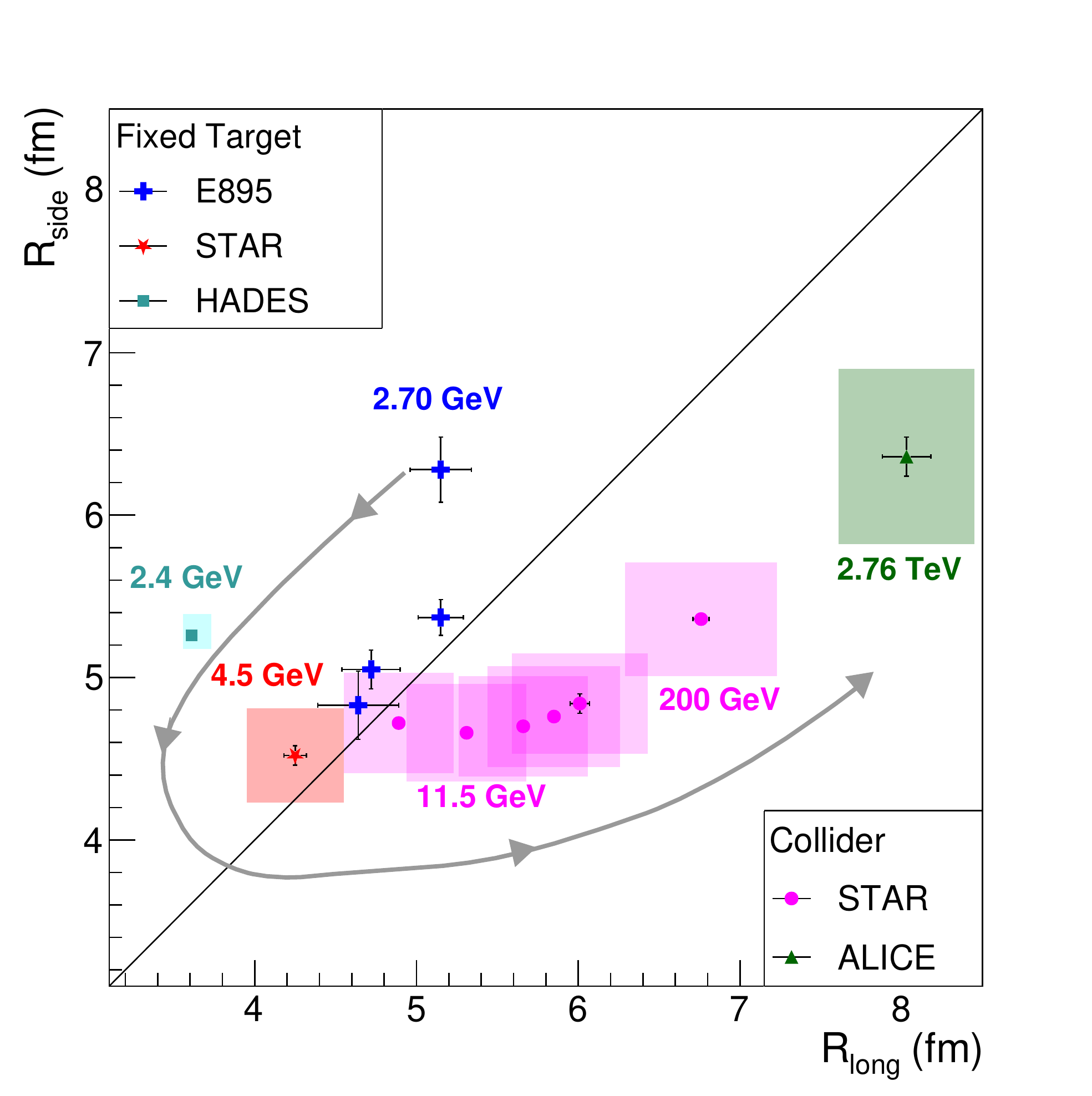}
\caption{
    \Rside vs. \Rlong, which measures the prolateness/oblateness of the pion emitting source when viewed from beside the beam.
    HADES~\cite{Adamczewski-Musch:2019dlf}, ALICE~\cite{Aamodt:2011mr} and STAR~\cite{Adamczyk:2014mxp} points include systematic errors; 
    E895~\cite{PhysRevLett.84.2798} 
    show statistical errors only. STAR fixed target data correspond to pion pairs with $\left\langle k_T\right\rangle=0.22$~GeV/c from 0-5\% centrality events.
    The various centrality, \pt, and $k_\text{T}$ cuts used in the different experiments are discussed in the text.
    The grey curve indicates the evolution of the shape, as the
    collision energy is increased.
}
\label{fig:fig23}
\end{figure}

Figure \ref{fig:fig19} shows fits of the form in Eq. \ref{eq:CFTheoryDefinition} (red lines) to the experimental correlation function defined in Eq. \ref{eq:CFExpDefinition} (blue stars).
The three panels show projections of the correlation function onto the $q_\text{out}$, $q_\text{side}$, and $q_\text{long}$ axes.
Data here are for $\pi^- \pi^-$ pairs created from tracks with transverse momentum $0.1 < \pt < 0.3$ GeV/$\textit{c}$, from events in the 0-10\% centrality range.
The transverse momentum of the pairs is required to be in the range $ 0.15 < k_\text{T} < 0.6 $ GeV/$\textit{c}$.
These cuts are chosen to match as closely as possible those in the E895 experiment, which used the same \pt cuts and corresponded to approximately 0-11\% centrality \cite{PhysRevLett.84.2798}.
There is a slight suppression at $q_\text{side} \approx 0$ and $q_\text{long} \approx 0$ due to the Coulomb repulsion of like-sign pion pairs.
The three-dimensional fit reproduces the data reasonably well.
The correlation functions are
fit via maximum likelihood~\cite{Lisa:2005dd}, but chi-square is often used as a measure of the fit.
For the fit in Fig.~\ref{fig:fig19}, $\chi^2/{\rm ndf}=122272/108811=1.12$.
Fits discussed here have $\chi^2/{\rm ndf}=\text{1 -- 2}$.
While not perfect, these reasonable fits can be used to extract radii that characterize
the spacetime extent of the source.

\subsubsection{Comparison with published data from similar energies}

Figure \ref{fig:fig20} shows the excitation function of the three femtoscopic radii for the
HADES~\cite{Adamczewski-Musch:2019dlf}, E895~\cite{PhysRevLett.84.2798}, STAR, and E866~\cite{Ahle:2002mi}
experiments.
The comparison with data from E866 is complicated by several issues. 
Firstly, a different centrality definition was employed, and it is unclear
how to translate this into the more commonly-used characterization of
the fraction of the inelastic cross section.  
Secondly, the narrow spectrometer acceptance of E866
 did not cover midrapidity (it covered $-0.30\lesssim y\lesssim -0.05$)
and has a 
higher transverse momentum lower limit.
Thirdly, unlike the other results to which we compare (and most other
measurements), the \mt-dependent analysis was not performed
in the LCMS. Nevertheless, the E866 results with the closest event and track selection criteria to the present results are included for context. 
The E895 and E866 points show a monotonically decreasing beam energy dependence.
The fixed-target STAR points are consistent with this trend within the uncertainties.
Femtoscopic radii reported~\cite{Adamczewski-Musch:2019dlf} by the HADES collaboration
  are clearly in quantitative disagreement with the trends observed in Fig.~\ref{fig:fig20}; we discuss
  this further below.

The \Rside radius primarily reflects the spatial extent of the pion emitting source, whereas \Rout convolves this with the emission duration of the fireball \cite{Mount:2010ey,PhysRevC.70.044907,RISCHKE1996479}.  
Figure \ref{fig:fig21} shows the radii as functions of the transverse mass 
$\mt = \sqrt{m_{\pi}^2 + k_\text{T}^2}$,
where $m_{\pi}$ is the pion mass.
In order to match analysis cuts from the E866 data, here the STAR points use a wider transverse momentum cut of $0.15 < \pt < 0.8$ GeV/$\textit{c}$, and include events from the 0-15\% centrality range.
The decrease in \Rside and \Rout with increasing \mt has been attributed to transverse flow, and the decrease in \Rlong is attributed to longitudinal flow \cite{PhysRevC.70.044907,Makhlin1988}.
High-\mt pairs come from smaller regions within the source and do not reflect the system's overall size \cite{Adamczyk:2014mxp}.
The STAR points agree very well with those from E895 and E866 for \Rside and \Rlong, as well as for \Rout at high \mt.
For \Rout the STAR points are slightly below E895 and E866 at low \mt, 
but agree within uncertainties.

Figure \ref{fig:fig22} shows the centrality dependence of the radii.
Here we combine $\pi^+\pi^+$ and $\pi^-\pi^-$ pairs and use the wider transverse momentum range of $0.15 < \pt < 0.8$ GeV/$\textit{c}$.
The radii decrease for more peripheral events due to the smaller geometric size of the initial participant region and the subsequent emission region at freezeout.

\subsubsection{Evolution from oblate to prolate freezeout configuration}

Figure \ref{fig:fig23} shows \Rside vs. \Rlong for several different data sets.
STAR FXT and BES points use low-$k_\text{T}$, $\pi^+ \pi^+$ and $\pi^- \pi^-$ pion pairs, with $\langle k_\text{T} \rangle \approx $ 0.22 GeV/c.
Events are drawn from the 0-5\% centrality range.
The ALICE point also corresponds to 0-5\% centrality, but a slightly higher $\langle k_\text{T} \rangle$ of  $\approx $ 0.26 GeV/c.
The E895 points use the cuts discussed above.
The collision energies ($\sqrt{s_{NN}}$) corresponding to each experiment are indicated in GeV.
The significantly different
acceptance and use of a different frame by E866~\cite{Ahle:2002mi} affects the longitudinal
radius in a way very different from that for the sideward.
Hence, it makes little
sense to include E866 data in a graph which plots \Rside versus \Rlong;
it is not shown in Fig.~\ref{fig:fig23}, which is a direct comparison
of similar measurements over three orders of magnitude in energy.

A clear evolution in the freezeout shape is indicated in the figure.
Lower energy collisions generally produce more oblate systems, and the shape of the emission region tends to become more prolate as the collision energy is increased.
In this representation, the evolution follows a ``swoosh'' systematic,
  indicated by the grey curve drawn to guide the eye.
This trend reflects the
evolution from stopping-dominated dynamics at low collision energies,
to the approximately longitudinally-boost-invariant scenario at the highest
energies.
The STAR fixed-target point has $\Rside\approx\Rlong\approx4.5$~fm, indicating a source that is approximately round when viewed from the side, just at the transition point between oblate and prolate geometry.

\subsubsection{Comparison to generic expectations due to a first-order phase transition at RHIC}

The femtoscopic radii reported~\cite{Adamczewski-Musch:2019dlf} by the
  HADES collaboration are consistent with the oblate shape reported by E895 at low energy.
However, it is clear from Figs.~\ref{fig:fig20} and \ref{fig:fig23} that the HADES radii
  are considerably smaller than would be expected by simple extrapolation of earlier data.
The reasons for this are unclear, and speculation is outside the scope of this paper.
However, there are several experimental systematic effects that can shift femtoscopic
  radii.
These include
  treatment of Coulomb effects,
  non-Gaussian shapes of the underlying correlation function
  (probed by varying the fitting range in $|\vec{q}|$),
  and $\vec{q}$-dependent particle-identification purity.
In addition, collision centrality definition and single-particle acceptance can vary slightly
  from one experiment to the next, complicating comparisons.
Ideally, such effects would be corrected for, or accounted for as part of the systematic uncertainty;
  however, subtle effects may persist and may be unique to a given experimental configuration.
Importantly, however, most of these effects affect $R_{\rm out}$,
  $R_{\rm side}$ and $R_{\rm long}$ in the same way.
Differences and (especially) ratios of femtoscopic radii are less susceptible to experiment-specific
  artifacts.

In the absence of collective flow, the emission timescale is
  related~\cite{PhysRevD.33.1314} to the
  transverse femtoscopic radii as $\beta^2\tau^2=R_{\rm out}^2-R_{\rm side}^2$,
  where $\beta$ is the transverse velocity of the emitted pions.
While collective flow complicates the interpretation~\cite{PhysRevC.70.044907}, 
  an extended emission
  timescale will increase $R_{\rm out}$ relative to $R_{\rm side}$.
A long emission timescale may arise if the system equilibrates close to
  the deconfinement phase boundary and then evolves through a first-order
  phase transition in the QCD phase
  diagram~\cite{PhysRevD.33.1314,PhysRevC.37.1896,PhysRevD.33.1314}.
Relativistic hydrodynamic calculations~\cite{RISCHKE1996479} 
  predict that a QCD first-order phase
  transition should produce a peak in the energy dependence of 
  $R_{\rm out}/R_{\rm side}$ near the QGP creation threshold.
Such a peak has also been suggested~\cite{Lacey:2014wqa,Lacey:2015yxg} as a signal of hadronization
  near a critical end point in the QCD phase diagram.

\begin{figure}
    \centering
    \includegraphics[width=0.48\textwidth]{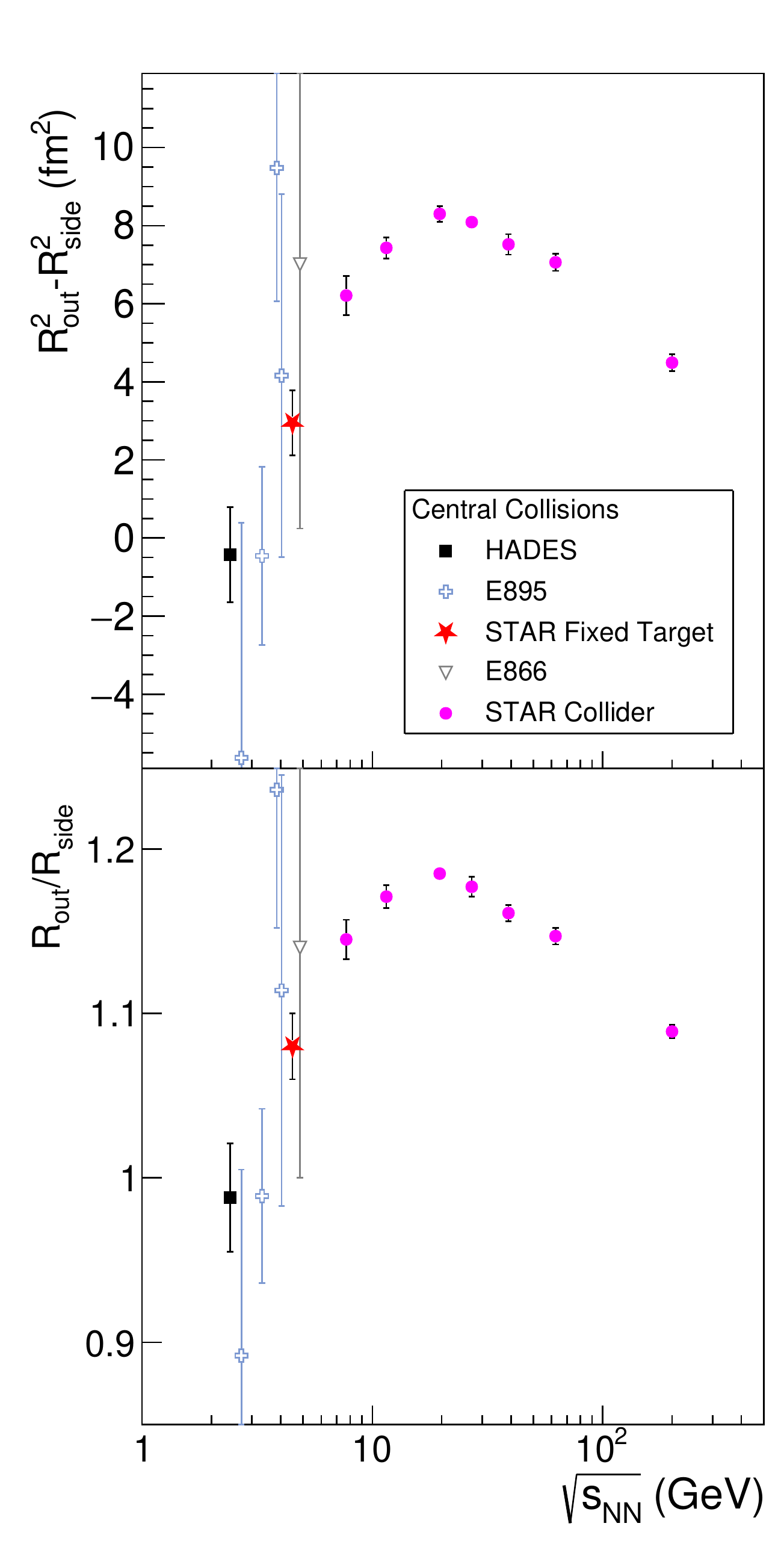}
    \caption{Top: The difference between the squared transverse femtoscopic radii are
    plotted as a function of the collision energy for central collisions.
    Bottom: The energy dependence of the ratio of the transverse radii.
    The centrality and $m_T$ values for the high-statistics datasets are:
    HADES~\cite{Adamczewski-Musch:2019dlf} (0-10\%, 349~MeV/c);
    STAR Fixed Target (this work) (0-10\%, 303~MeV/c);
    STAR Collider~\cite{Adamczyk:2014mxp} (0-10\%, 326~MeV/c).
    The values for the earlier measurements are:
    E895~\cite{Lisa:2000hw} (0-11\%, 330~MeV/c); 
    E866~\cite{Ahle:2002mi} (0-15\%, 295~MeV/c).
    Only statistical errors are indicated, as
    changing the centrality~\cite{Lisa:2005dd} or transverse mass selection
    slightly will affect $R_{\rm out}$ and $R_{\rm side}$ similarly; see
    the text for a discussion of systematic effects, which can shift
    STAR datapoints, together, by $\sim5\%$ ($\sim20\%$) for $R_{\rm out}/R_{\rm side}$
    ($R^2_{\rm out}-R^2_{\rm side}$).
 }
    \label{fig:RoutRside}
\end{figure}

The energy dependences of $R_{\rm out}^2-R_{\rm side}^2$ and 
  $R_{\rm out}/R_{\rm side}$ are shown in 
  Fig.~\ref{fig:RoutRside}.
Both quantities exhibit a clear peak at
  $\sqrt{s_{NN}}\approx20$~GeV, an interesting energy where other 
  observables~\cite{Adamczyk:2014mzf,Adamczyk:2014fia,Adamczyk:2014ipa,Adamczyk:2013dal,Adam:2020unf} show nontrivial trends with energy.
The earlier E895 and E866 results are consistent with the trend from STAR
  and HADES, but their statistical uncertainties are much too large to
  resolve a peak of the magnitude observed.
Systematic errors on these quantities are given in Table~\ref{tab:RoutRsideSystematicErrors}
  for STAR measurements, both in collider and fixed-target modes.
Importantly, the systematic errors are common for all STAR points (collider and fixed-target), hence
  variations in (for example) the treatment of Coulomb effects will move all data points similarly, not
  changing the peak structure.

\begin{table}[]
    \centering
    \begin{tabular}{l|cc}
    source                                                      & $\delta\left(\frac{R_{\rm out}}{R_{\rm side}}\right)$ & $\delta\left(R_{\rm out}^2-R_{\rm side}^2\right)$ \\
    \hline
    variation in centrality                                     & 1\%       & 8\% \\
    50~MeV/c variation in $\left\langle m_T \right\rangle$      & 2\%       & 8\% \\
    varying fit range in $|\vec{q}|$                            & $<1\%$    & 10\% \\
    varying track-merging cut                                   & 4\%       & 10\% \\
    treatment of Coulomb effects                                & 1\%       & 6\%
    \end{tabular}
    \caption{Systematic error estimates for the quantities plotted in figure~\ref{fig:RoutRside}.
    First row considers using a 2-12\% selection rather than a 0-10\% selection.
    Track-merging cuts, fit-range systematics and Coulomb effects are discussed in~\cite{Adams:2004yc,Adamczyk:2014mxp}.
    \label{tab:RoutRsideSystematicErrors}}
\end{table}

First measurements of $R_{\rm out}^2-R_{\rm side}^2$ and $R_{\rm out}/R_{\rm side}$
  at the highest energies at RHIC~\cite{Adler:2001zd,Adams:2004yc} were similar to
  values measured at lower energies, contrary to 
  some expectations of a long lifetime~\cite{Heinz:2002un,Lisa:2005dd}.
This ``puzzle''~\cite{Heinz:2002un} was eventually partly understood
  as arising from
  a number of independent complications that tend to reduce the extended lifetime 
  signal~\cite{Pratt:2009hu}.
Figure~\ref{fig:RoutRside} suggests two other reasons that the signal was not
  observed.
Firstly, the energy of collisions at full RHIC energy ($\sqrt{s_{NN}}=200$~GeV) 
  may be too high above the threshold energy for QGP formation; at such high
  energies, the extended lifetime signal is predicted to disappear~\cite{RISCHKE1996479}.
Secondly, the early femtoscopic data from E895 and E866 was insufficiently
  precise to discern the peak revealed by higher-statistics data.
The STAR low energy measurements address this second issue.
Indeed, the entire STAR fixed-target program is crucial for identifying such
  energy-dependent trends.


\section{Summary}

In this first set of results from fixed-target running at the STAR experiment, we report that the directed flow ($v_1$) of protons and $\Lambda$ baryons is in line with existing systematics at higher and lower energy. This is important, as the directed flow of baryons shows a sign change and a minimum just above the present beam energy, while the directed flow of net baryons shows a double sign change \cite{Adamczyk:2014ipa, Adamczyk:2017nxg}. This is one of the most intriguing experimental results from the BES-I program, as well as one of the most difficult for models to explain \cite{Steinheimer:2014pfa, Konchakovski:2014gda, Ivanov:2014ioa, Nara:2016hbg, Singha:2016mna, Nara:2017qcg, Nara:2020ztb}. 

We have also presented the first measurements of azimuthal anisotropy of charged pions and neutral kaons at these energies.  Both show directed flow ($v_1$) signals in the direction opposite to that of the baryons, continuing trends observed at higher energies. The difference between $\pi^+$ and $\pi^-$ flow becomes stronger as the collision energy is reduced, perhaps signaling isospin or Coulomb dynamics. Interestingly, within the relatively large statistical uncertainties, the data are consistent with constituent quark scaling of elliptic flow, an effect proposed at much higher energies to arise from quark coalescence in the QGP phase.

Femtoscopic radii with charged pions are consistent with earlier measurements of energy, transverse mass, and centrality systematics.
Collisions at $\sqrt{s_{\rm NN}}=4.5$~GeV are in the transition region between dynamics dominated by
  stopping (producing an oblate source) and boost-invariant dynamics (prolate source).

More importantly, these new measurements with much-improved statistics, together with recent HADES results, reveal 
a long-sought peak structure that may be caused by the system evolving through a first-order
  phase transition from the QGP to the hadronic phase.
Previous results were insufficiently precise to detect this effect.
This is the promise of an experimental program that revisits heavy ion collisions in this energy range:
  improving the quantitative precision of measurements, with well-understood systematics consistent over
  a broad energy range, may produce qualitatively new opportunities.
Now that the predicted~\cite{RISCHKE1996479} $R_{\rm out}/R_{\rm side}$ energy systematic has been
  revealed, it deserves theoretical attention from hydro and transport modelers.
The magnitude and width of the structure may allow an estimate of the latent heat of the QCD deconfinement transition.

Overall, while these measurements are important and of interest on their own, they also pave the way for the FXT energy scan with nominally one hundred times more events at each of 9 beam energy points.  The FXT energy scan is an integral part of the BES-II program at RHIC which began in early 2019. It extends the reach of the STAR experiment across an important energy regime of high baryon chemical potential, 
ranging from 420 to 720 MeV \cite{Cleymans:2005xv}, corresponding to collision energies from $\sqrt{s_{\text{NN}}} = 7.7$ GeV down to 3.0 GeV.

\section{Acknowledgements}

We acknowledge valuable discussions with Yasushi Nara and Horst St\"ocker. We thank the RHIC Operations Group and RCF at BNL, the NERSC Center at LBNL, and the Open Science Grid consortium for providing resources and support.  This work was supported in part by the Office of Nuclear Physics within the U.S. DOE Office of Science, the U.S. National Science Foundation, the Ministry of Education and Science of the Russian Federation, National Natural Science Foundation of China, Chinese Academy of Science, the Ministry of Science and Technology of China and the Chinese Ministry of Education, the Higher Education Sprout Project by Ministry of Education at NCKU, the National Research Foundation of Korea, Czech Science Foundation and Ministry of Education, Youth and Sports of the Czech Republic, Hungarian National Research, Development and Innovation Office, New National Excellency Programme of the Hungarian Ministry of Human Capacities, Department of Atomic Energy and Department of Science and Technology of the Government of India, the National Science Centre of Poland, the Ministry  of Science, Education and Sports of the Republic of Croatia, RosAtom of Russia and German Bundesministerium fur Bildung, Wissenschaft, Forschung and Technologie (BMBF), Helmholtz Association, Ministry of Education, Culture, Sports, Science, and Technology (MEXT) and Japan Society for the Promotion of Science (JSPS).



\include{FXT_AuAu_bib}
\bibliography{FXT_AuAu_bib}{}
\end{document}